\documentclass[showpacs,showkeys,amsmath,amssymb,floatfix]{revtex4}
\usepackage{amsfonts}
\usepackage{amsmath}
\usepackage{amssymb}
\usepackage{graphicx}
 \usepackage{hyperref}

\newcommand{\wb}{\omega_{\mathrm{b}}}
\newcommand{\wo}{\omega_{\mathrm{o}}}

\newcommand{\wL}{\omega_{\mathrm{L}}}
\newcommand{\m}{m^*}
\newcommand{\nv}{n_{\mathrm{v}}}
\newcommand{\doublefig}{.5\textwidth}
\newcommand{\singlefig}{.7\textwidth}

\begin{document}
\title{Classical and quantum nonlinear localized excitations in discrete systems}
\author{FR Romero} \email{romero@us.es}
\author{JFR Archilla}
\author{F Palmero}
\author{B S\'{a}nchez-Rey}
\author{A \'{A}lvarez}
\author{J Cuevas}
\author{JM Romero}
\affiliation {Grupo de F\'{\i}sica No Lineal. Departamento de
F\'{\i}sica at\'{\o}mica, molecular y nuclear y Departamento de
F\'{\i}sica Aplicada I. Universidad de Sevilla.
    Avda. Reina Mercedes, s/n. 41012-Sevilla (Spain)}

\date{April 10, 2005 }



\begin{abstract}
Discrete breathers, or intrinsic localized modes, are spatially
localized, time--periodic, nonlinear excitations that can exist
and propagate in systems of coupled dynamical units. Recently,
some experiments show the sighting of a form of discrete breather
that exist at the atomic scale in a magnetic solid. Other
observations of breathers refer to systems such as
Josephson--junction arrays, photonic crystals and
optical-switching waveguide arrays. All these observations
underscore their importance in physical phenomena at all scales.
The authors review some of their latest theoretical contributions
in the field of classical and quantum breathers, with possible
applications to these widely different physical systems and to
many other such as DNA, proteins, quantum dots, quantum computing,
etc.

\end{abstract}

\maketitle

\section{Introduction}

Nonlinear physics of discrete systems has attained an enormous
development in the previous years. A landmark is the prove of the
existence of spatially localized, time--periodics oscillations in
discrete nonlinear systems, by MacKay and Aubry ~\cite{MA94}. They
are now termed discrete breathers (DB), nonlinear localized
excitations, or intrinsic localized modes, and ever since there
are many experimental data and theoretical results published about
these important new phenomenon in physics. They appear in lattices
of oscillators interacting through nonlinear forces. The central
class of system for which the concept of DB was developed is
classical, autonomous spatially discrete Hamiltonian or reversible
systems. A type of systems of this class is called Klein--Gordon
lattices. They are Hamiltonian systems consisting of one degree of
freedom anharmonic Hamiltonian oscillators coupled weakly in a
lattice, that is, the frequency of oscillation of each oscillator
varies non--trivially with the amplitude. Most studies on
breathers have been done utilizing this type of models, with
applications to biomolecules, such as DNA, solids, etc. Other
classes of systems where breathers occur are: autonomous forced
damped systems, and time--periodically forced systems. Considering
applications to molecular crystals, quantum effects must be taken
into account and the concept of quantum discrete breather should
be developed. This review summarize some of the latest theoretical
contributions in the field of localized nonlinear excitations
carried out by the Nonlinear Physics Group of the Sevilla
University. The structure of this work has the following scheme:

\begin{itemize}
\item Section II: Breathers in Klein--Gordon lattices with competing
short and long--range interactions
\begin{itemize}
\item A: DNA model with dipole--dipole coupling
\item B: Numerical methods for obtaining breathers and determining
their stability
\item C: Bifurcations
\end{itemize}
\item Section III: Moving breathers in homogeneous DNA chains
\begin{itemize}
\item A: Moving breathers in DNA chains without curvature
\item B: Moving breathers in bent DNA chains
\item C: Parameters values of DNA
\end{itemize}
\item Section IV: Interaction of moving breathers with mass
impurities
\item Section V: Interaction of moving breathers with vacancies
\item Section VI: Small and large amplitude breathers in
Fermi-Pasta-Ulam lattices
\item Section VII: Dark breathers in Klein--Gordon lattices
\item Section VIII: Polarons in biomolecules
\item Section IX: Quantum breathers
\begin{itemize}
\item A: Quantum breathers in a translational invariant lattice
\item B: Trapping in lattices with broken translational
symmetry
\end{itemize}
\item Section X: Thermal evolution of enzyme-created oscillating
bubbles in DNA
\end{itemize}

\section{Breathers in Klein--Gordon lattices with competing short and long--range
interactions}

Most of the studies about breathers in Klein--Gordon lattices
refer to models which include only short--range interacting forces
between the oscillators. This section reviews some of the results
relative to breathers solutions in Klein--Gordon chains  which
include also long--range interactions.

 A one--dimensional
Klein--Gordon lattice is a chain of identical oscillators with an
on--site potential $V$, each one of them coupled with its nearest
neighbours through a parameter $C\ll 1$. The Hamiltonian of a
chain of $N$ oscillators is:

\begin{equation}\label{ham}
    H=\sum_{n=1}^N\left(\frac{1}{2}\dot
    u_n^2+V(u_n)+\frac{1}{2}C(u_n-u_{n-1})^2\right),
\end{equation}
where $u_n$ represents the coordinates of the oscillators referred
to their equilibrium positions, and $V(u_n)$ represents the
on--site potential.

 In addition to the short--range interaction coupling, other types of
 forces can also exist. For example, if each oscillator has an
 intrinsic dipole moment, there exist also long--range forces due
 to the dipole--dipole coupling, and consequently the Hamiltonian
 must be modified. The existence of these two competing forces could
 alter substantially the properties of the breathers
 solutions. The study is presented in the next subsections,
 selecting a DNA model which has also into
 account the existence of dipole--dipole coupling.

\subsection{ DNA model with dipole--dipole coupling }
In DNA there exists different kinds of interactions between the
main atomic groups. One of them is the stacking interaction
between neighbour bases along the DNA axis, these are short--range
forces which stabilize the DNA structure and hold one base over
the next one forming a stack of bases. There exist also
long--range forces, due to the finite dipole moments of the
hydrogen bonds within the nucleotides ~\cite{GMCR97,MCGJR99}. The
dipole--dipole interactions between different base pairs become
critical when the geometry of the double strand of DNA is taken
into account, as the distance between base pairs and, therefore,
the intensity of the coupling between them, depends on the shape
of the molecule ~\cite{ACMG01,ACG01,CPAR02}.

The Peyrard-Bishop DNA model ~\cite{PB89} was initially introduced
considering only the stacking interactions between base pairs. It
can be modified by adding an energy term that takes into account
the dipole--dipole forces. The Hamiltonian can be written as:

\begin{equation}\label{eq:ham_straight}
    H=\sum_{n=1}^N\left(\frac{1}{2}m\dot u_n^2+V(u_n)+
    \frac{1}{2}C(u_{n+1}-u_n)^2+
    \frac{1}{2}\sum_{p\ne n}\frac{J}{|p|}u_{n+p} u_n\right).
\end{equation}
The term $\frac{1}{2}m\dot u_n^2$ represents the kinetic energy of
the nucleotide of mass $m$ at  the $n$th site of the chain, and
$u_n$ is the variable representing the transverse stretching of
the hydrogen bond connecting the bases. The on--site potential is
the Morse potential, i.e., $D(e^{-bu_n}-1)^2$, represents the
interaction energy due to the hydrogen bonds within the base
pairs, where $D$ is the well depth and represents the dissociation
energy of a base pair, and $b^{-1}$ is related to the width of the
well. The stacking energy is $\frac{1}{2}C(u_{n+1}-u_n)^2$, where
$C$ is the stacking coupling constant ~\cite{PB89} . The last term
of the Hamiltonian is the long-range dipole--dipole interaction
term, where $J=q^2/4\pi\varepsilon_0d^3$ is the dipole long-range
interaction coupling constant ~\cite{CAGR02}, with $q$ being the
charge transfer due to the formation of the hydrogen bonds, and
$d$ the distance between base pairs, which is supposed to be
constant. It is possible to consider $D=1/2$, $b=1$, $m=1$ without
loss of generality.

\subsection{Numerical methods for obtaining breathers
and determining their stability}

A static breather (SB) can be obtained by solving the full
dynamical equations, which can be obtained applying Hamilton's
equations to (\ref{eq:ham_straight}). This aim can be achieved
using common methods based on the anticontinuous limit
~\cite{MA94,MA96,M97,AMM99}. The implementation of these methods
basically consists in calculating the orbit of an isolated
oscillator at fixed frequency $\wb$, and using this solution as a
seed to solve the complete dynamical equations by means of a
Newton--Raphson continuation method. A SB can be obtained for a
given values of the dipole--dipole coupling parameter, $J$, and
the stacking coupling parameter, $C$. This solution can be
continued by varying $C$ and maintaining $J$ constant or vice
versa.

The linear stability of a breather can be studied by performing a
Floquet analysis ~\cite{A97}. To this end, it is necessary to
consider the evolution of a perturbation $\xi$ through the
dynamical equations:

\begin{equation}\label{eq:perturb}
    \ddot{\xi_n}+V''(u_n)\xi_n+
    C(2\xi_n-\xi_{n+1}-\xi_{n-1})+
    \sum_{p\ne n}\frac{J}{p}\xi_{n+p}=0 .
\end{equation}
The functions $\Omega(0)\equiv (\{\xi_n(0)\},\{\dot{\xi}_n(0)\})$
must be integrated until $t=T$, giving $\Omega(T)\equiv
(\{\xi_n(T)\},\{\dot{\xi}_n(T)\})$. There exists a matrix of
dimension $2N$ called the \emph{monodromy}, which is defined by
$\Omega(T)=\mathcal{M}\Omega(0)$.

The spectrum of this matrix has the following property: if
$\lambda$ is an eigenvalue, then $\lambda^*$, $1/\lambda$ and
$1/\lambda^*$ are also eigenvalues. Therefore, stability implies
that all the eigenvalues have modulus unity ~\cite{Tesis}.

The procedure of continuation of SBs can be done as long as an
integer multiple of their frequency does not resonate with any of
the frequencies of the linear modes (or phonons). This resonance
provides an analytical expression for the upper boundary
delimiting the existence of SBs: $C=(\wb/\wo)^2-1/4+3\zeta(3)J/8$.

\subsection{Bifurcations}

The study of the bifurcation loci of SBs is interesting for three
different reasons: a)~It gives the range of existence of the SBs;
b)~It gives the regions in the parameter space where the breathers
are stable (and, therefore, physically observable in real systems)
or unstable; c)~Finally, but of paramount importance, it gives the
values of the parameters where the breathers are movable.

Two types of bifurcations appear as $C$ and $J$ are varying:
1)~Stability bifurcations: they occur when a breather changes its
stability. The existence of this type of bifurcation is a
necessary condition for the existence of movable breathers (see
Section \ref{sec:moving}). 2)~Breather extinctions: they occur
when a breather is not continuable any longer.

When only stacking interaction is considered, only stability
bifurcations occur and moving breathers are possible. However,
with only dipole-dipole interactions there are no stability
bifurcations and, therefore, moving breathers do not exist.

When both interactions exist simultaneously, the analysis of
1-site (site-centered) and 2-site (bond-centered) breathers,
permits to establish the following staments: A) 1--site breathers
can be moved only for $J<J_{c1}$. B) 2--site breathers are movable
for $J<J_{c2}$ being $J_{c2}>J_{c1}$. They are also movable for $J
\in [J_{c2},J_o)$. Therefore, $J_o$ establishes the maximum value
of $J$ (and also the minimum value of $C$) for which breathers can
be movable, i.e., \emph{there are no moving breathers for $C<J_o$
and for $J>J_o$}. This is the reason why $J_o$ is called the
\emph{mobility limit} ~\cite{CAGR02}.

\begin{figure}
    \begin{center}
        \begin{tabular}{cc}
            (a) & (b) \\
            \includegraphics[clip=true,width=\doublefig]{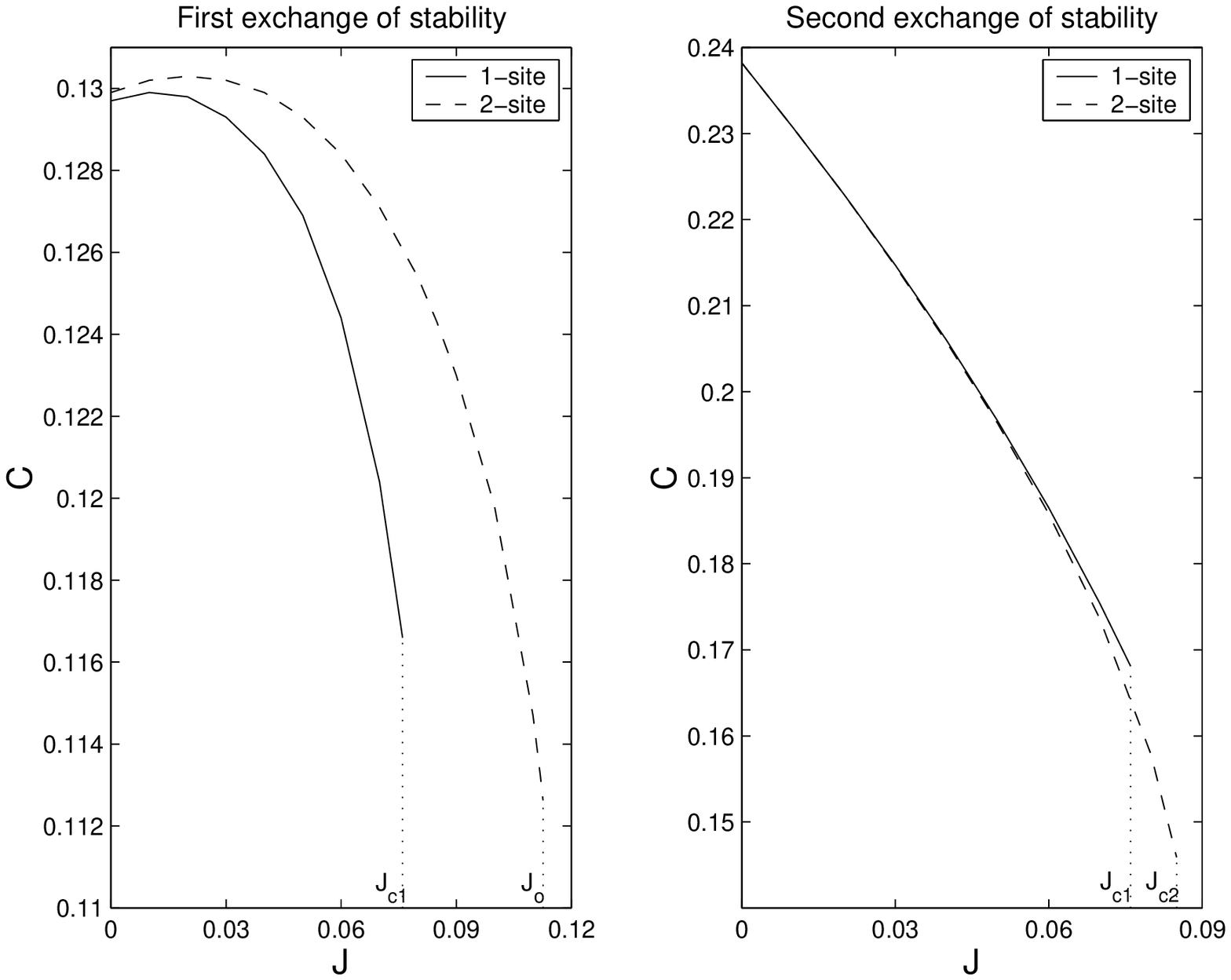} &
            \includegraphics[clip=true,width=\doublefig]{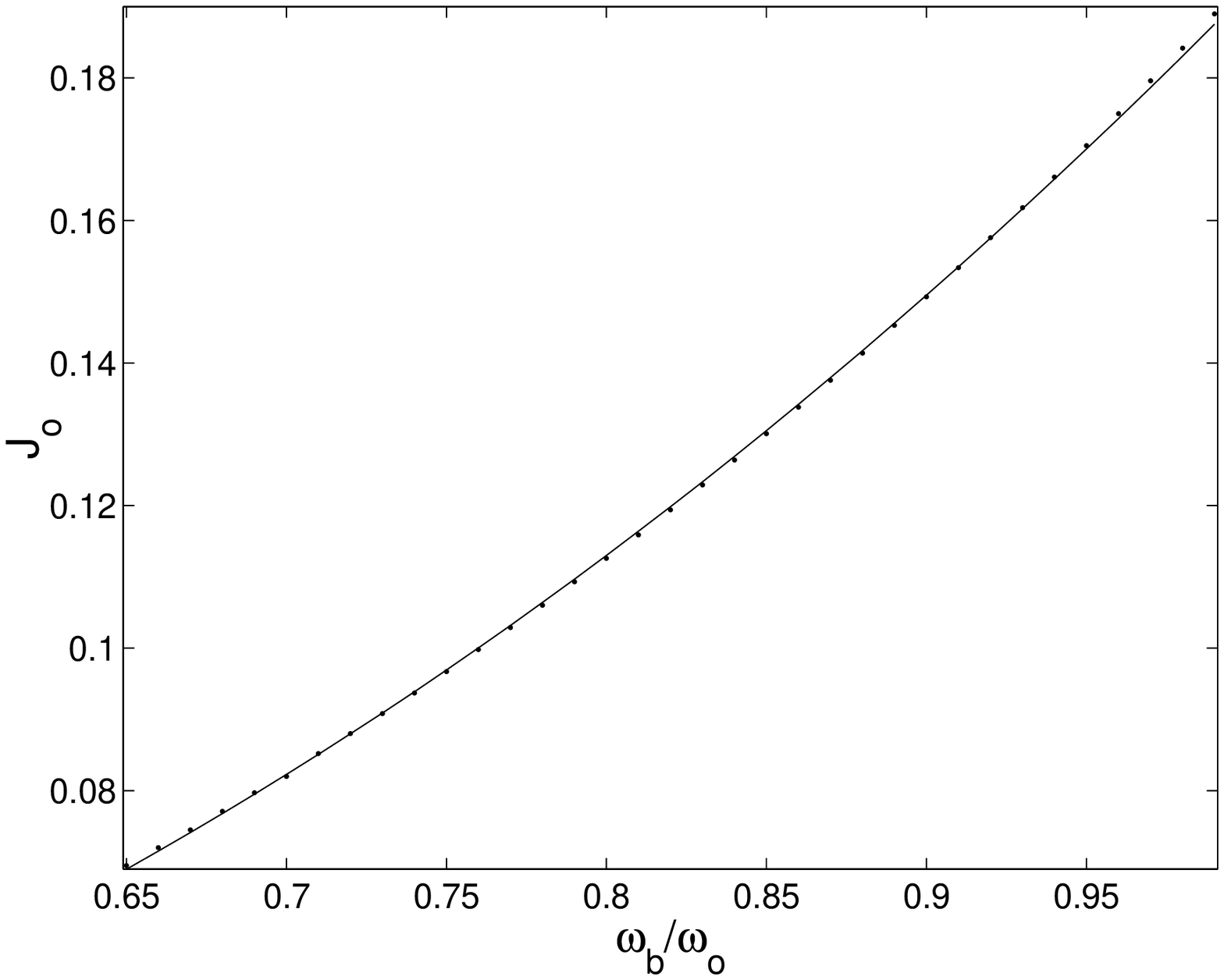} \\
        \end{tabular}
    \end{center}
    \caption{(a) Bifurcation loci for the first and the second exchange
    of stability in 1-site and 2-site breathers. The values corresponding to
    $\wb=0.8$ are: $J_{c1}=0.076$, $J_{c2}=0.085$, $J_o=0.11263$.
    (b) Dependence of $J_o$ with respect to the breather frequency.}
    \label{Fig1}
\end{figure}

The bifurcation loci for the stability bifurcations are shown in
Fig. \ref{Fig1}(a). Also, in Fig. \ref{Fig1}(b) the dependence of
the mobility limit with respect to the breather frequency is
shown. This curve fits very well to the relation $J_o=A\wb^r$,
where A=$0.1921\pm0.0002$ and r=$2.377\pm0.005$.


\section{Moving breathers in homogeneous DNA chains}
\label{sec:moving}

This section is dedicated to the study of travelling breathers, or
moving breathers, in DNA chains described by the Peyrard--Bishop
modified model. The first subsection presents this study for
straight line DNA chains, or chains without curvature. The next
subsection considers DNA chains which have been bent and adapted
to a parabola.

\subsection{Moving breathers in DNA chains without curvature}

\begin{figure}
    \begin{center}
            \includegraphics[clip=true,width=\singlefig]{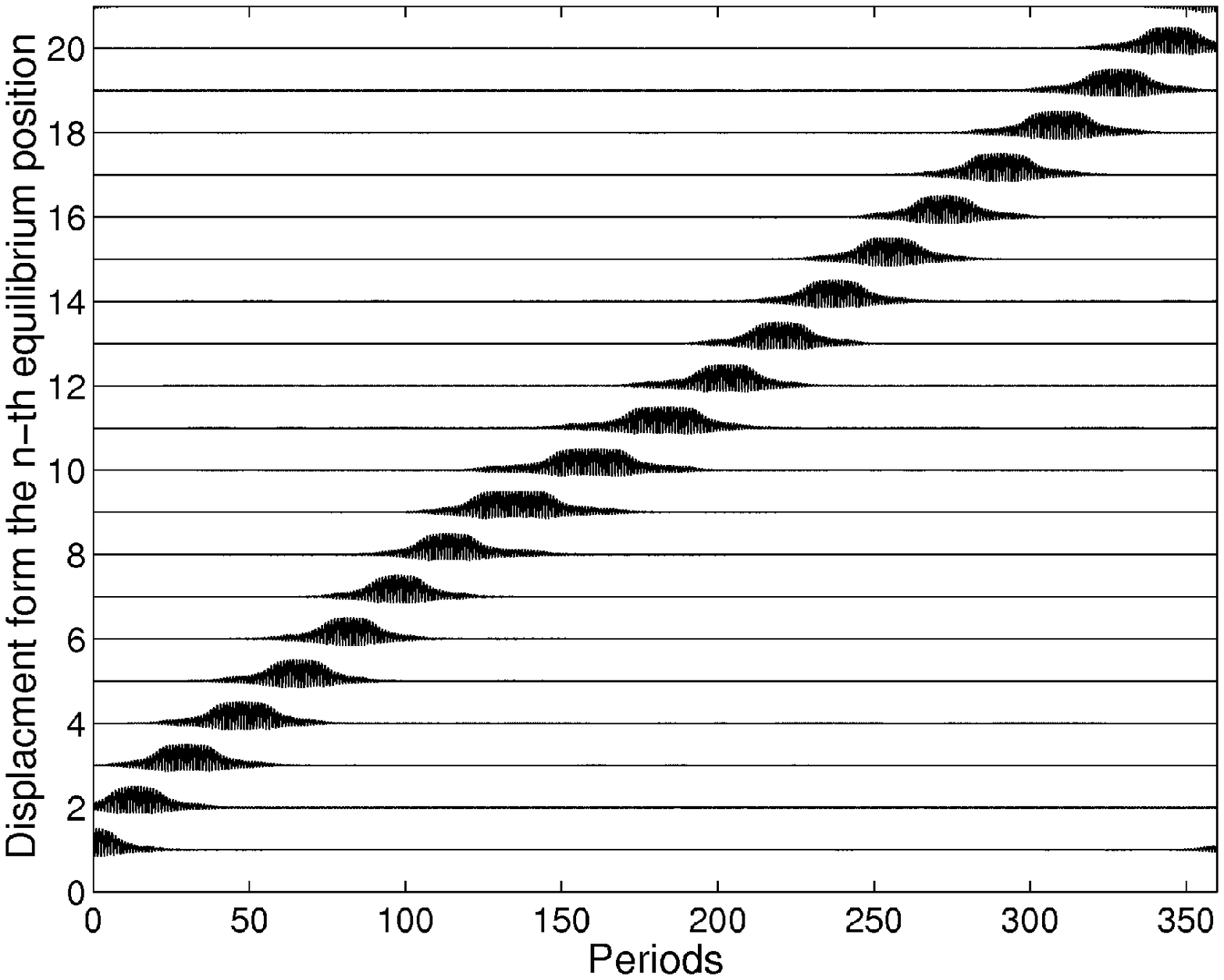}
    \end{center}
    \caption{Evolution of a moving breather for $C=0.1244$, $J=0.06$ and
    an initial `kick' $\lambda=0.10$. The breather involves basically three
    sites that oscillate in phase.}
    \label{Fig2}
\end{figure}

Contrary to static breathers, moving breathers are not exact
solutions of the dynamical equations of the model and they have a
finite life due to the effects of phonon radiation. A moving
breather is obtained by `kicking' a static breather, i.e.,
perturbing its velocity components. However, not all breathers can
be moved, and not every kick can move a breather. Therefore, it is
important to know the conditions that a breather must fulfill in
order to be movable and what the characteristics of the
perturbations are. As indicated in ~\cite{CAT96,AC98,Cretegny},
the following steps must be performed in order to obtain a moving
breather: 1)~To look for the existence of the two complementary
stability bifurcations for the 1--site breather and the 2--site
breather. Their bifurcation loci must have a region where they are
fairly close. The static breather that can be moved should be
obtained for values of the parameters close enough to these
bifurcation loci. However, as shown in ~\cite{CAGR02} it is only
necessary that the static breather is ``quasi-stable'', i.e., not
very stable or not very unstable; 2)~To perturb the breather with
the velocity components of the marginal mode at the neighbouring
bifurcation. Time evolution of a moving breather is shown in Fig.
\ref{Fig2}.


An useful concept used for describing the breather dynamics is its
effective mass ~\cite{CAT96,AC98}. It is a measure of the breather
inertia to external forces. The breather velocity must be
perturbed in a direction colinear to the \emph{marginal mode} in
order to obtain moving breathers. This mode is the responsible for
the stability bifurcation. If the normalized marginal mode is
$\mathbf{V}$, the perturbation added to the breather velocities,
which are zero at $t=0$, is given by $\lambda \mathbf{V}$, and
$\lambda$ is the magnitude of the perturbation. Thus, the kinetic
energy added to the breather by the initial kick is
$1/2\lambda^2$. It is found that the resulting translational
velocity of the breather, $v$, is proportional to $\lambda$. Then,
the concept of effective mass can be defined through the relation
$m^{*}v^2/2=\lambda^2/2$. Therefore, $m^*=(\lambda/v)^2$.
Consequently, moving breathers can be considered as a
quasi-particle with mass $m^*$. The effective mass is a
quantitative measure of the breather mobility. Larger mass
indicates smaller mobility.

\begin{figure}
    \begin{center}
            \includegraphics[clip=true,width=\singlefig]{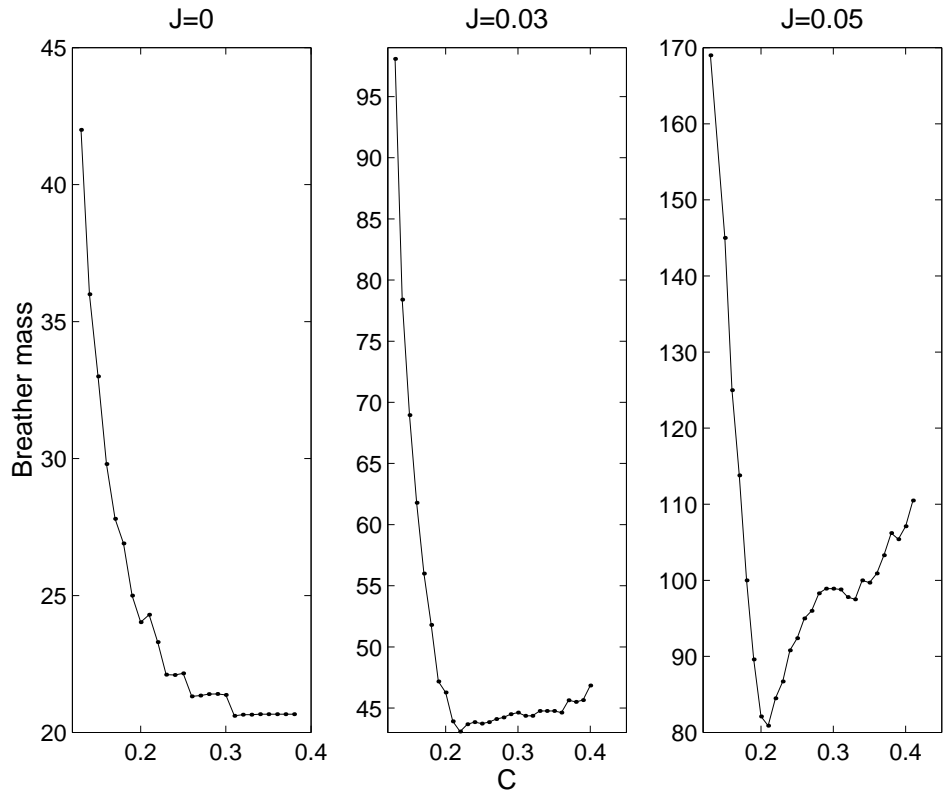}
    \end{center}
    \caption{Breather mass versus stacking coupling parameter $C$ for
    dipole--dipole coupling parameter $J=0$ (left), $J=0.03$ (center) and
    $J=0.05$ (right). Note that the maximum value of the mass occurs at
    the first bifurcation, and that the breather mass for $J=0.05$ is always greater
    than the mass for $J=0$. Note also the different scales at the Y-axis.}
    \label{Fig3}
\end{figure}

Numerical studies show that, for high values of the dipole--dipole
coupling, the predicted range of existence of moving breathers
decreases and, in some cases, it is impossible for the breather to
be moved even in the neighbourhood of the stability bifurcations.
Another important result is that the dipole--dipole interaction
affects to the mobility of the breathers. The study of the
dependence of the breather effective mass, shows that the mobility
decreases when the intensity of the dipole--dipole coupling
increases (see Fig. \ref{Fig3}).


\subsection{Moving breathers in bent DNA chains}

Suppose a DNA chain lying in a plane which has been bent and
adapted to a parabola of curvature $\kappa$ (that is, the location
of the n-th base pair is determined by the equation $y_n=\kappa
x_n^2/2$) ~\cite{ACG01}. All the dipole moments are perpendicular
to the plane and parallel among them. Thus, the modified
Hamiltonian of Eq. (\ref{eq:ham_straight}) is:

\begin{equation}\label{eq:ham_bent}
    H=\sum_{n=1}^N\left(\frac{1}{2}m\dot u_n^2+V(u_n)+
    \frac{1}{2}C(u_{n+1}-u_n)^2+
    \frac{1}{2}\sum_{p\ne n}\frac{J}{|\vec r_n-\vec r_p|^3}u_{n+p} u_n\right),
\end{equation}
where $\vec r_n=(x_n,y_n)$, and it is assumed that the chain is
inextensible, so that the distance between neighbouring sites
remains constant: $|\vec r_n-\vec r_{n+1}|\equiv d$.

\begin{figure}
  \begin{center}
        \begin{tabular}{cc}
            (a) & (b) \\
            \includegraphics[clip=true,width=\doublefig]{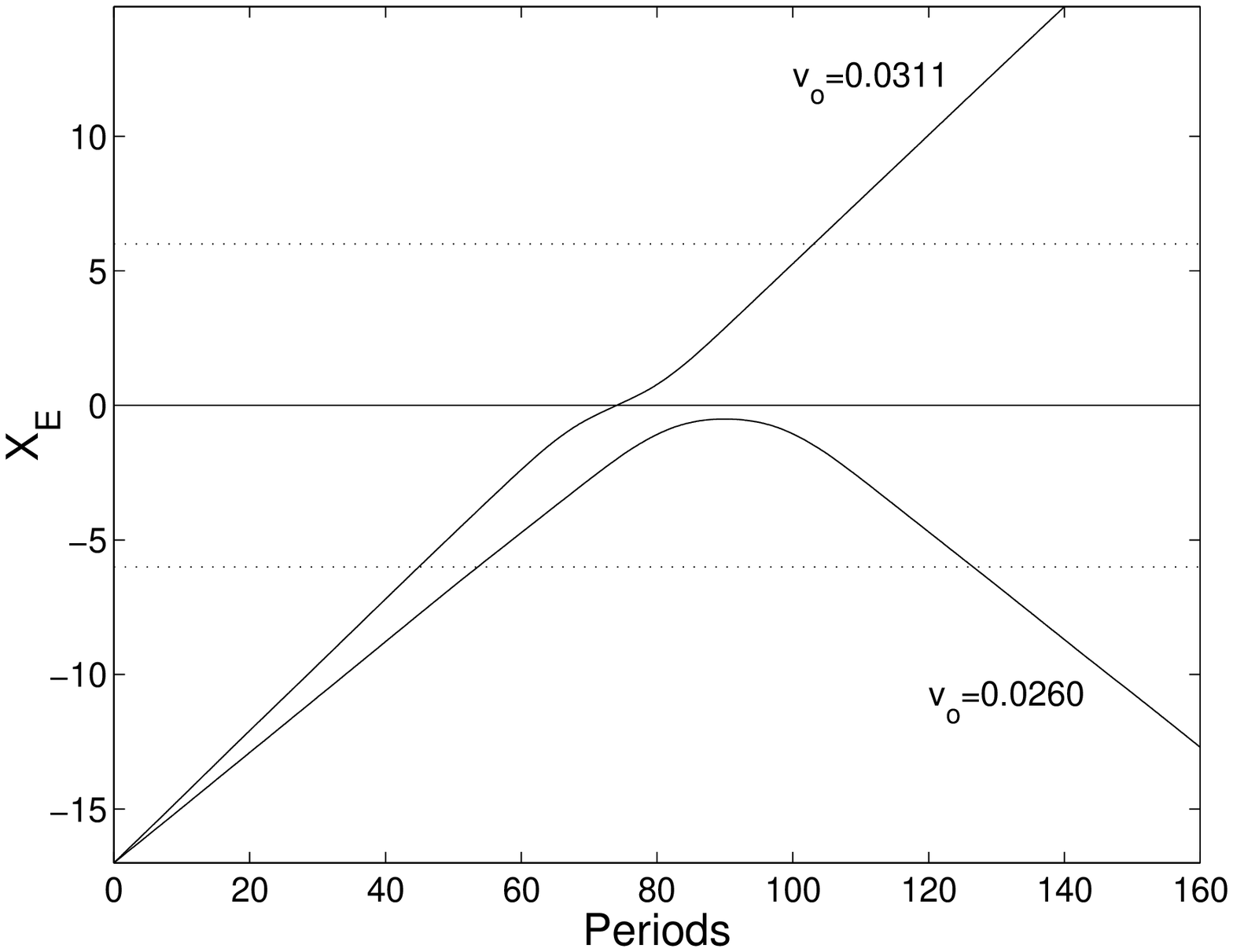} &
            \includegraphics[clip=true,width=\doublefig]{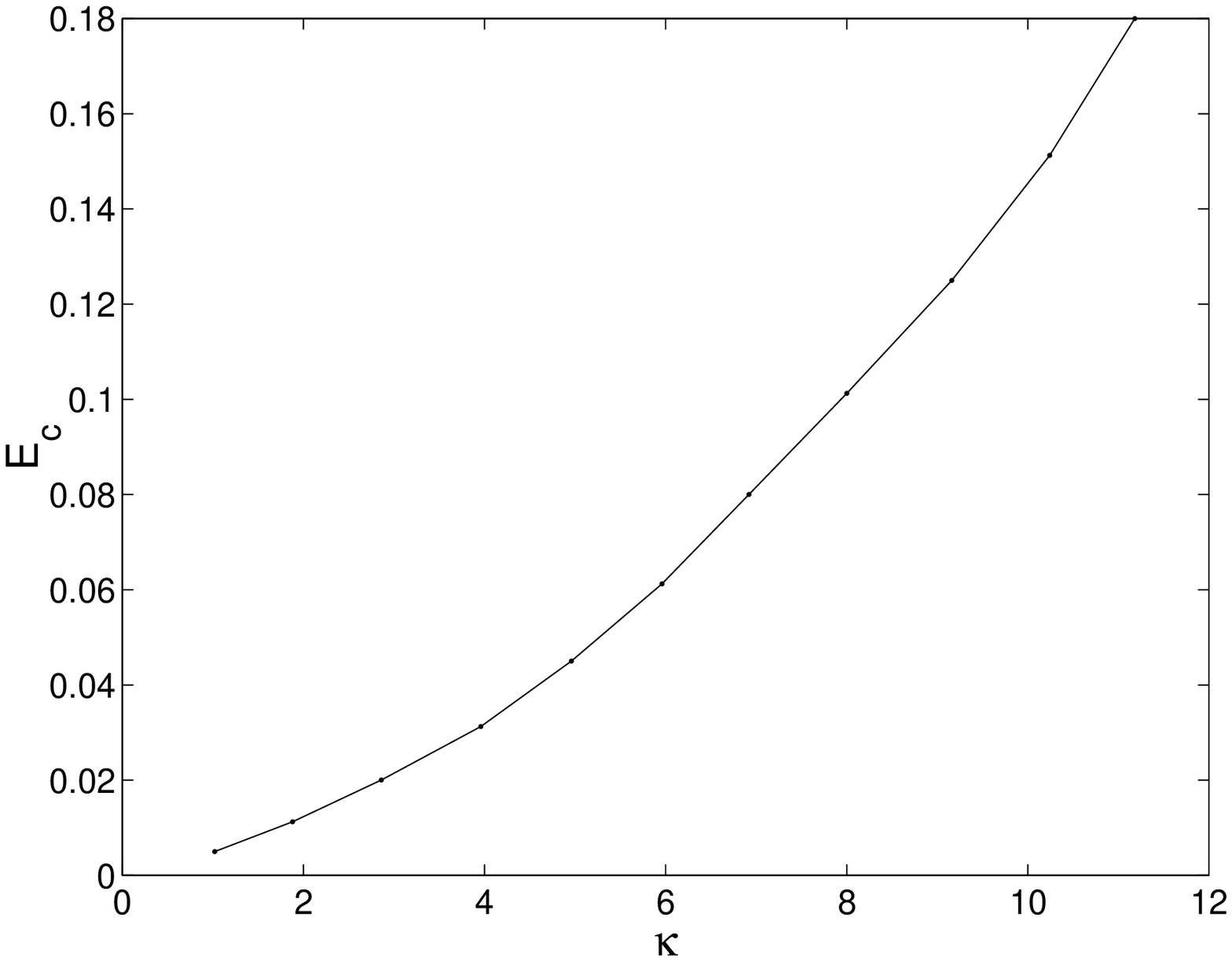} \\
        \end{tabular}
  \end{center}
  \caption{(a)Evolution of the breather energy centre ($X_E$) for two
  different initial velocities. The curvature is $\kappa=2$.
  (b) Critical kinetic energy $E_c$ versus curvature $\kappa$.}
  \label{Fig4}
\end{figure}

 Fig. \ref{Fig4}(a) illustrates the evolution of the energy
centre ~\cite{CAGR02} of a moving breather in a bent chain. If the
added kinetic energy, $E=\lambda^2/2$, is smaller than a critical
value $E_c$, the breather rebounds, but, if $E>E_c$, the breather
passes through the bending point. Fig. \ref{Fig4}(b) shows that
the critical energy increases monotonically with the curvature.


\begin{figure}[t]
  \begin{center}
    \includegraphics[clip=true,width=\singlefig]{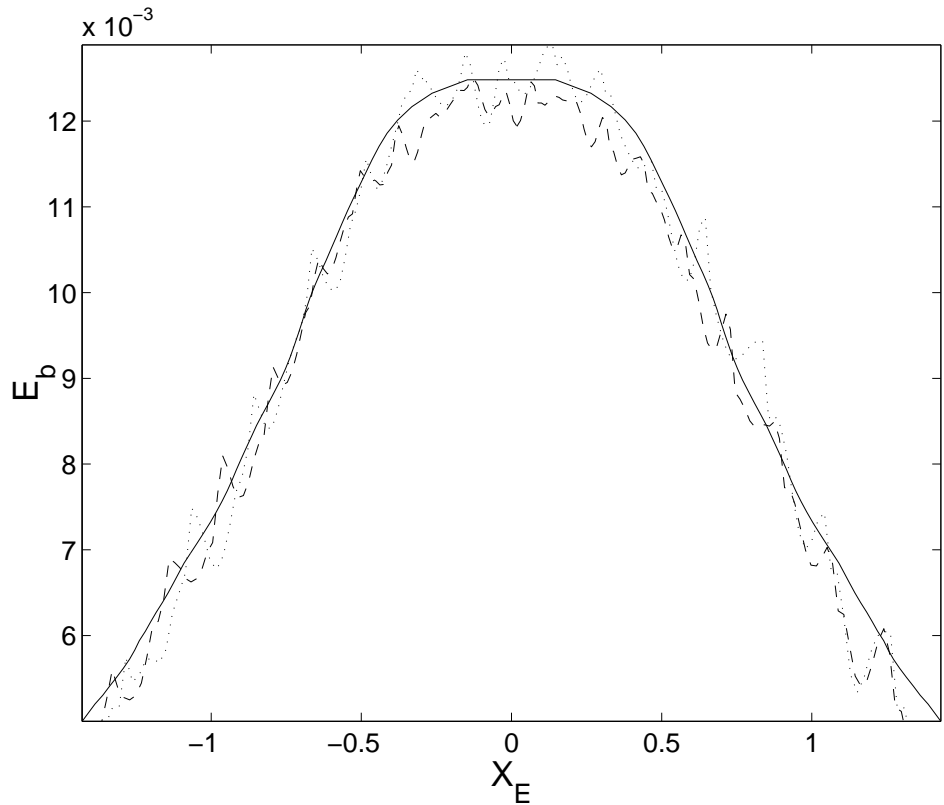}
  \end{center}
  \caption{Potential barrier calculated finding the turning points
  (solid line) and using Eq. (\ref{barrier}) for
  $E=0.0162$ (dashed lined) and $E=0.0200$ (dotted line).
  The curvature is $\kappa=2$ and the critical energy
  is $E_c\approx0.0126$. The zero value of the energy centre
  ($X_E$) represents to the bending point.}
  \label{Fig5}
\end{figure}

Chain bending acts as a hindrance for the movement of breathers.
This hindrance resembles to the experimented by a particle moving
in a potential barrier. In this case, breather can be consider as
a quasi-particle and this barrier can be calculated by finding the
points where breathers rebound, i.e., the turning points for
different values of $E$. Furthermore, if the breather has a
constant effective mass $\m$, this potential barrier can be
obtained using the expression:

\begin{equation} \label{barrier}
    E_b(x)=\frac{1}{2}\lambda^2[1-(v(x)/v_o)^2],
\end{equation}
where $v(x)$ is the translational velocity and $v_o$ is its value
at $t=0$. Fig. \ref{Fig5} show that there exists a good agreement
between the barriers calculated using both methods for a given
value of $\kappa$. The barrier calculated by the second method
exhibits an irregular shape, whose origin lies in the non-uniform
behaviour of the translational velocity due to the discreteness of
the system ~\cite{Cretegny}. This result confirms that, in this
case, a moving breather behaves as a particle of constant
effective mass $\m$.


It is worth remarking that a similar study has been performed in
bent alpha--helix proteins ~\cite{AGCC02}.

\subsection{Parameters values of DNA}

The results presented are valid for all chains with both short
range and long range interactions given by equations
(\ref{eq:ham_straight}) and (\ref{eq:ham_bent}). When they model a
DNA chain, the following parameter values should be used
~\cite{CSAH04}: $D=0.04$~eV, $b=4.45$~\AA$^{-1}$, $d=3.4$~\AA,
$J=0.0031$~eV$/$\AA$^2$, $C=(0.01,10)$~eV$/$\AA$^2$, $m=300$~amu
and the time unit is $0.88$~ps. For this set of values, the
minimal kinetic energy for a moving breather to cross the bending
point is small, and, in consequence, it seems plausible that most
MBs will have enough energy to pass through bending points,
although the ones with low energy will be reflected.

\section{Interaction of moving breathers with mass impurities}

Sometimes a chain can have a single point inhomogeneity, that is,
the chain is homogeneous except at a single site where,for
example, an impurity or a different type of particle is located.
Moving breathers can be generated far apart the impurity and be
launched towards it. This section presents some results relative
to the effects of the collisions of moving breathers with an
impurity in a Klein--Gordon chain.

Consider a Klein--Gordon chain with nearest neighbours attractive
interactions, with  Hamiltonian given by:

\begin{equation}\label{ham2}
    H=\sum_{n=1}^N\left(\frac{1}{2}\dot
    u_n^2+V_n(u_n)+\frac{1}{2}C(u_n-u_{n-1})^2\right),
\end{equation}
where $V_n(u_n)=D_n(e^{-u_n}-1)^2$ is a substrate potential at the
n-th site. The inhomogeneity is introduced assuming a different
well depth at a single site, i.e.,
$D_n=D_o(1+\alpha\delta_{n,0})$, where $\alpha\in[-1,\infty)$ is a
parameter which tunes the magnitude of the inhomogeneity. The
particle located at $n=0$ is an impurity.

\begin{figure}
\begin{tabular}{cc}
    (a) & (b) \\
    \includegraphics[clip=true,width=\doublefig]{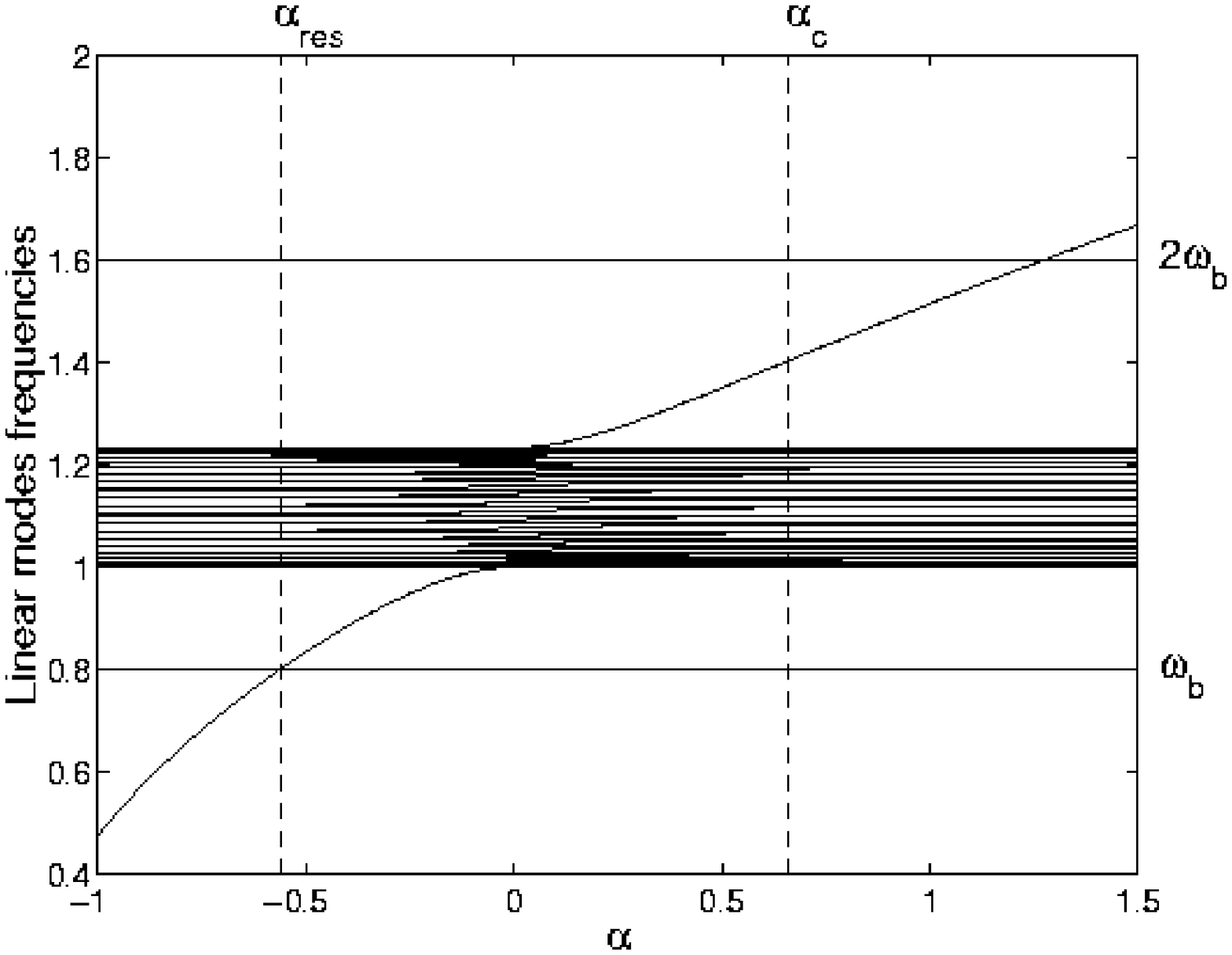} &
    \includegraphics[clip=true,width=\doublefig]{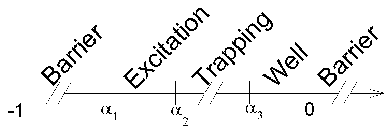}
\end{tabular}
\caption{(a) Frequencies of the linear modes versus the parameter
$\alpha$. At $\alpha=\alpha_{res}$ and $\alpha=\alpha_c$, two
different bifurcations occur, being the first one due to the
resonance between the impurity mode and the breather. (b)
Different regimes in the interaction of a moving breather with an
impurity.} \label{Fig6}
\end{figure}

The dynamical equations can be linearized if the amplitudes of the
oscillations are small enough. These equations have $N-1$
non-localized solutions (\emph{linear extended modes}) and one
localized solution, (\emph{linear impurity mode}). Their
frequencies, $\omega_E$ and $\omega_L$, are given, respectively,
by:
\begin{equation}\label{LEMs}
    \omega(q,\alpha)=\sqrt{\omega_o^2+4C\sin^2\frac{q(\alpha)}{2}},
    \quad
    \omega^2_{L}=\wo^2+2C+\mathrm{sign}(\alpha)\sqrt{\alpha^2\wo^4+4C^2},
\end{equation}
where $q\in(0,\pi]$ if $\alpha<0$ and $q\in[0,\pi)$ if $\alpha>0$.
Fig. \ref{Fig6}(a) shows the dependence on $\alpha$.


\begin{figure}[t]
\begin{tabular}{cc}
    (a) & (b) \\
    \includegraphics[clip=true,width=\doublefig]{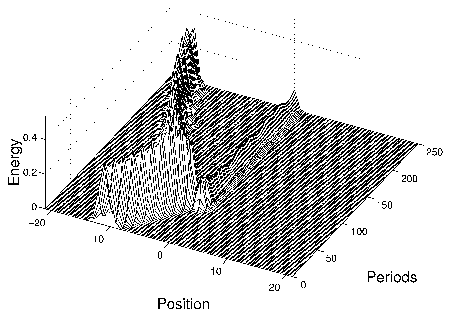} &
    \includegraphics[clip=true,width=\doublefig]{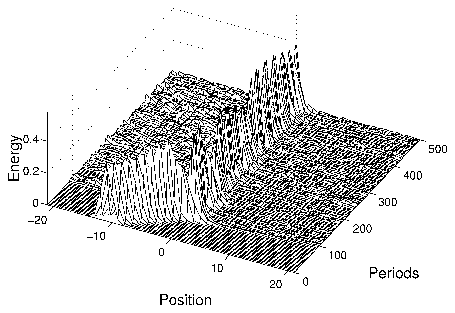}
\end{tabular}
\caption{(a) Interaction of a breather with an impurity for
$\alpha=-0.52$, which corresponds to the impurity excitation case.
(b) Evolution of the moving breather for $\alpha=-0.3$, which
corresponds to the trapping case. The moving breather becomes
trapped by the impurity; afterwards, the breather emits phonon
radiation and its energy centre oscillates between the sites
adjacent to the impurity.}\label{Fig7}
\end{figure}

Numerical simulations show that, depending of the critical values
of the parameter $\alpha$, there exist four different regimes in
the moving breather--impurity interaction ~\cite{CPAR02b}:
1)~\emph{Barrier}. The impurity acts as a potential barrier. It
occurs either with $\alpha>0$ or $\alpha\in(-1,\alpha_1)$ with
$\alpha_1<0$. If $\alpha\gtrsim0$, the breather can pass through
the impurity provided the translational velocity is high enough
~\cite{CPAR02}.~\emph{Excitation}. The impurity is excited and the
breather is reflected. This case appears for
$\alpha\in(\alpha_1,\alpha_2)$, an example can be seen in Fig.
\ref{Fig7}(a). 3)~\emph{Trapping}. The breather is trapped by the
impurity. It occurs in the interval
$\alpha\in(\alpha_2,\alpha_3)$. When the moving breather is close
to the impurity, it becomes trapped while its center oscillates
between the neighbouring sites, as Fig. \ref{Fig7}(b) shows.
4)~\emph{Well}. The impurity acts as a potential well. It occurs
for $\alpha\in(\alpha_3,0)$ and consists of an acceleration of the
breather as it approaches to the impurity, and a deceleration
after the breather has passed through the impurity.


It is observed that the breather bifurcates with the zero solution
at $\alpha=\alpha_{res}$. That is, for $\alpha$ smaller than this
value, no impurity breather exists. At $\alpha=\alpha_{res}$, the
frequency of the impurity mode coincides with the moving breather
frequency, i.e., in (\ref{LEMs}), $\wL=\wb$.

The scenario for the trapped breathers when $\alpha<0$ is the
following: the impurity mode has $q=0$, and also all the particles
of the impurity breather vibrate in phase; this vibration pattern
indicates that the impurity breather bifurcates from the impurity
mode and it will be the only localized mode that exists when the
impurity is excited for $\alpha>\alpha_{res}$. Thus, when the moving
breather reaches the impurity, it can excite the impurity mode. For
$\alpha<\alpha_{res}$, the moving breather is always reflected. In
addition, the impurity breather does not exist. Therefore, there
might be a connection between both facts, i.e., the existence of the
impurity breather seems to be a necessary condition in order to
obtain a trapped breather. If $\alpha>0$, the impurity mode has
$q=\pi$ but the impurity breather's sites vibrate again in phase,
that is, the impurity breather does not bifurcate from the impurity
mode. There are two different localized excitations: the tails of
the (linear) impurity mode and the impurity breather. Thus, if the
moving breather reaches the impurity site, it will excite these
localized excitations. Therefore, we conjecture that the existence
of both linear localized entities at the same time may be the reason
why the impurity is unable to trap the breather when $\alpha>0$.

All these facts allows the formulation of the following trapping
hypothesis: \emph{The existence of an impurity breather for a
given value of $\alpha$ is a necessary condition for the existence
of trapped breathers. However, if there exists an impurity mode
with a vibration pattern different from the impurity breather
one's, the trapped breather does not to exist.}

\section{Interaction of moving breathers with vacancies}

Another type of local inhomogeneity that can exists in a
homogeneous chain is a vacancy, that is, the absence of a particle
at a single site. A moving breather can be launched towards the
vacancy and the effects of the collision should be analyzed.

\begin{figure}
\begin{center}
    \includegraphics[clip=true,width=\singlefig]{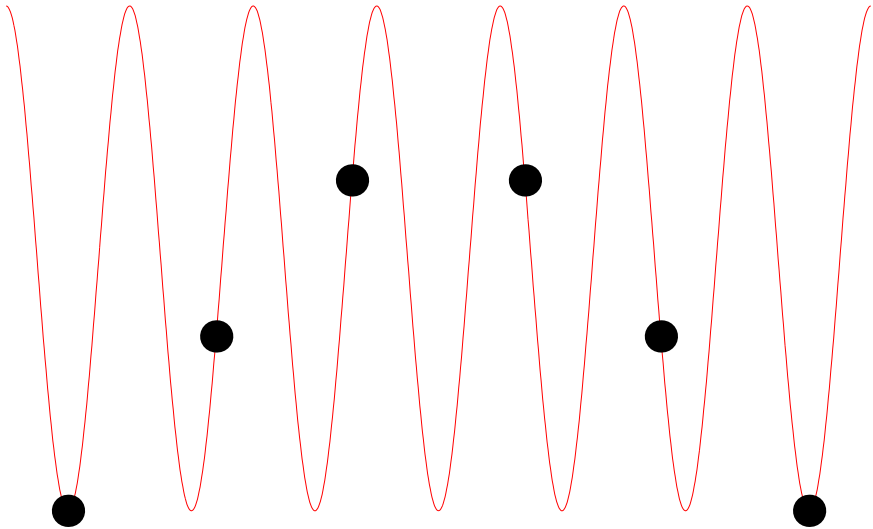}
\caption{Scheme of the Frenkel--Kontorova model with sine-Gordon
on--site potential, with full circles representing to the particles.}%
\label{Fig8}
\end{center}
\end{figure}

Consider a Hamiltonian Frenkel--Kontorova model with an anharmonic
interaction potential ~\cite{BK98}:

\begin{equation}
    H=\sum_n\frac{1}{2}\dot x_n^2+
    \frac{1}{4\pi^2}[1-\cos(2\pi x)]
    +C\,\frac{1}{2b^2}[\exp(-b(x_{n+1}-x_n-1))-1]^2,
\end{equation}
where $\{x_n\}$ are the absolute coordinates of the particles. The
choice of a periodic potential makes easy to represent a vacancy.
Thus, if the vacancy is located at site $\nv$ (see Fig.
\ref{Fig8}), the displacements of the particles with respect to
their equilibrium positions are:

\begin{equation}
\left\{ \begin{array}{ll} u_n=x_n-nL & n<\nv \\
\\ u_n=x_n-(n+1)L & n>\nv.\end{array} \right.
\end{equation}


 It is possible to generate a moving breather travelling towards the vacancy at
 the site $\nv$, and simulate the collision with this type of point inhomogeneity.

 The initial perturbation, $\{\mathbf V_n\}$ has been chosen as $\mathbf
V=\lambda(\ldots,0,-1/\sqrt{2},0,1/\sqrt{2},0,\ldots)$, where the
nonzero values correspond to the neighboring sites of the initial
center of the breather.

 When a moving breather reaches the site
occupied by the particle adjacent to the vacancy, i.e., the
location $\nv-1$, the particle can jump to the vacancy site or
remain at rest. If the former takes place, the vacancy moves
backwards. However, if the interaction potential is wide enough,
the particle at the $\nv+1$ site, can feel the effect of the
moving breather at the $\nv-1$ site and it can also move backwards
the vacancy site. In the last case, the vacancy moves forwards.

\begin{figure}
\begin{center}
\begin{tabular}{cc}
    (a) & (b) \\
    \includegraphics[clip=true,width=\doublefig]{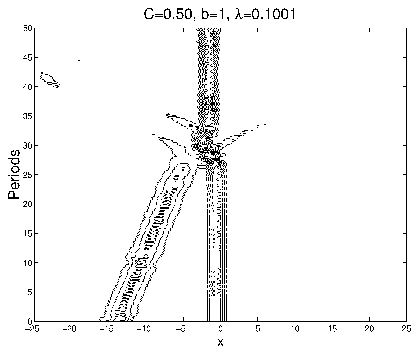} &
    \includegraphics[clip=true,width=\doublefig]{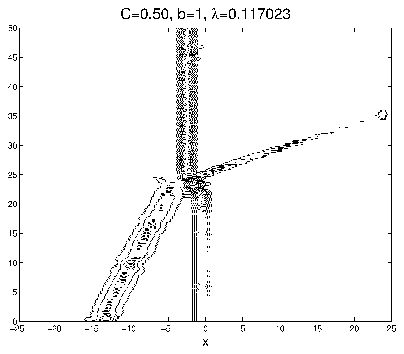}\\
    (c) & (d) \\
    \includegraphics[clip=true,width=\doublefig]{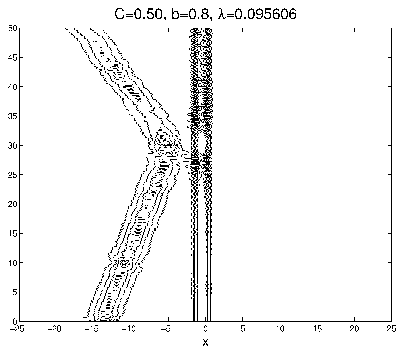} &
    \includegraphics[clip=true,width=\doublefig]{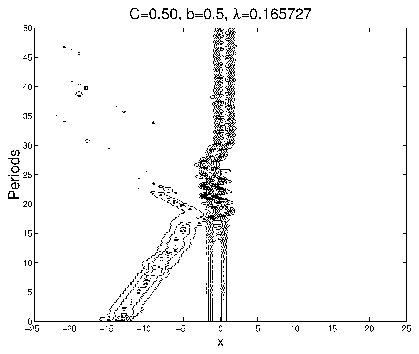}\\
\end{tabular}
\caption{Energy density plots of the interaction moving
breather--vacancy. The vacancy is located at $\nv=0$. (a) the
vacancy moves backwards; (b) the breather is transmitted by the
vacancy; (c) the breather is reflected and the
vacancy remains at rest; (d) the vacancy moves forwards.}%
\label{Fig9}
\end{center}
\end{figure}

Numerical simulations show that the occurrence of the three
different cases depends highly on the relative phase of the
incoming breather and the particles adjacent to the vacancy
~\cite{CKAER03}. However, some conclusions can be extracted: 1)
the incident breather always losses energy; 2) the breather can be
reflected, trapped (with emission of energy) or transmitted by the
vacancy, in analogy to the interaction moving breather-mass defect
~\cite{CPAR02b}; 3) the transmission of the breather (i.e. the
breather passes through the vacancy) can only take place if the
vacancy moves backwards, i.e. the particle to the left jumps one
site in the direction of the breather. An explanation of this fact
is that the particles to the right of the vacancy, in order to
support a moving breather, need a strong interaction which cannot
be provided by the interaction across a vacancy site, because the
distance correspond to the soft part of the Morse potential. Fig.
\ref{Fig9} illustrates these phenomena.


\begin{figure}
\begin{center}
\begin{tabular}{cc}
    \includegraphics[clip=true,width=\doublefig]{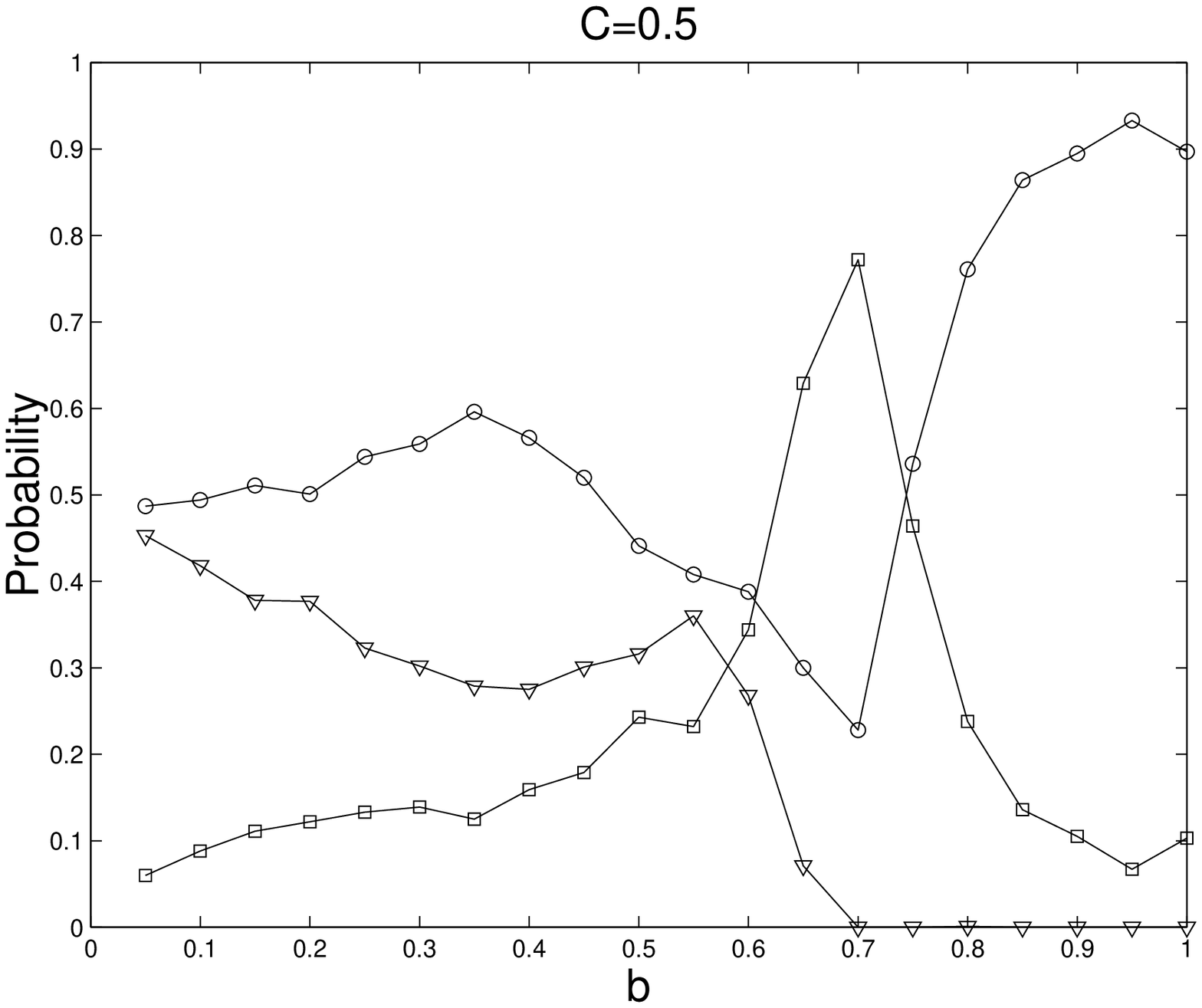} &
    \includegraphics[clip=true,width=\doublefig]{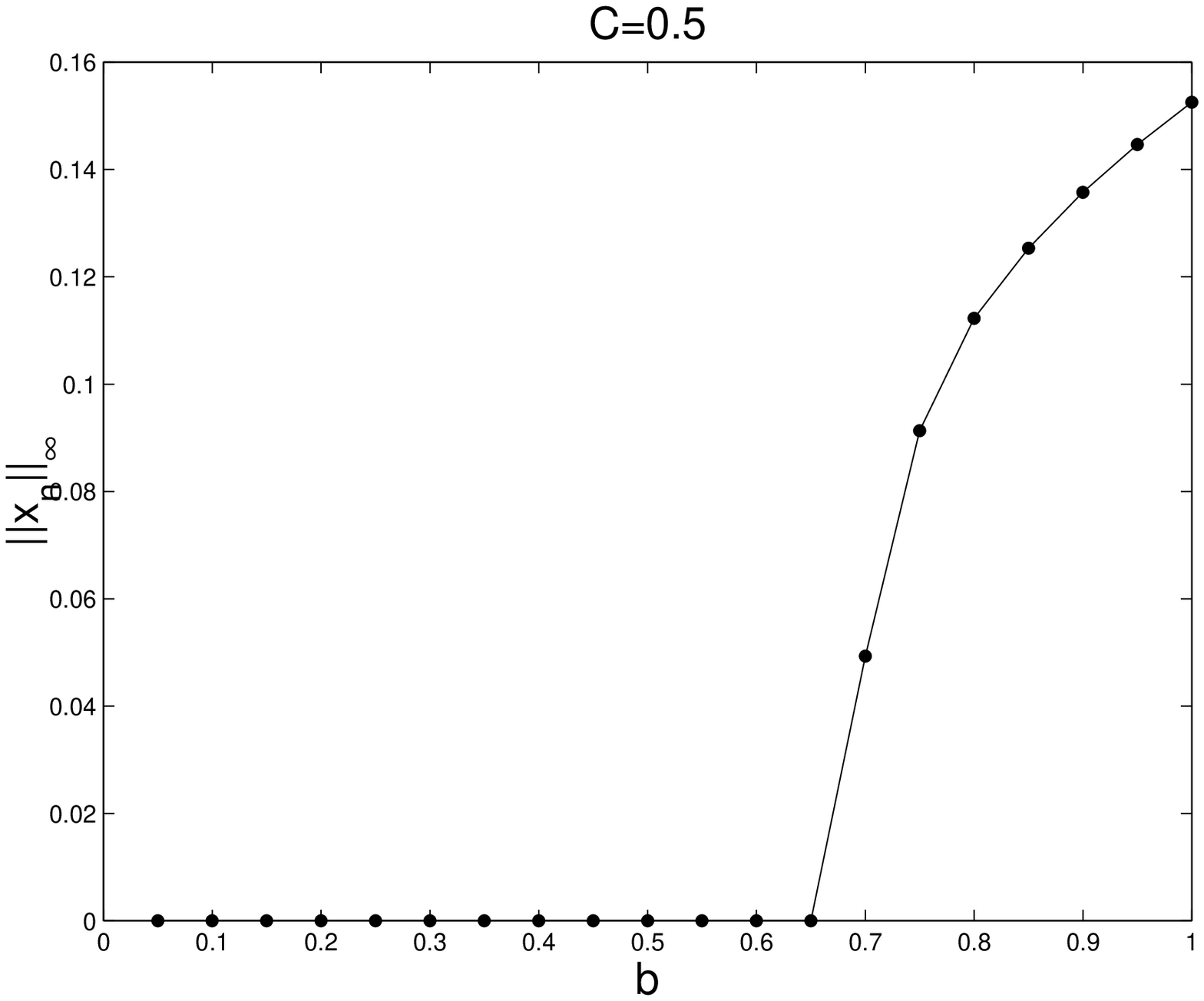}\\
\end{tabular}
\caption{(a) Probability that the vacancy remains at its site
(squares), moves backwards (circles) or moves forwards (triangles),
for a Gaussian distribution of $\lambda$ as a function of the
inverse potential width $b$. (b) Amplitude maxima of a vacancy
breather versus $b$. The vacancy breather disappears at the
bifurcation point. This is related to the vanishing of the forward
movement probability.}%
\label{Fig10}
\end{center}
\end{figure}

The moving breather--vacancy interaction is very sensitive to the
velocity of the breather and the distance between the breather and
the vacancy. Consequently, a systematic study of the state of the
moving breather and the vacancy after the interaction cannot be
performed. Therefore, we have performed a great number of
simulations each one consisting in launching a single breather
towards the vacancy site. In particular, we have chosen 1000
breathers following a Gaussian distribution of the perturbation
parameter $\lambda$ with mean value $0.13$ and variance $0.03$ for
different values of the parameters $b$ and $C$. Fig.
\ref{Fig10}(a) shows the probabilities that the vacancy remains at
its original site, or that it jumps backwards or forwards, for
$C=0.5$ and $C=0.4$. From this study, an important consequence can
be extracted: concerning to the parameter $b$, there exist two
different regions of values, separated by a critical value
$b_0(C)$, such that for $b>b_0(C)$, the probability that the
vacancy moves forwards is almost zero, whereas for $b<b_0(C)$,
this probability is significant.


The non-existence of forwards vacancy migration can be explained
through a bifurcation. These bifurcations are related to the
disappearance of the entities we call \emph{vacancy breathers}.
They are defined as breathers centered at the site neighboring to
the vacancy, e.g. the $\nv-1$ or $\nv+1$ sites. It can be observed
in Fig. \ref{Fig10}(b) that, for $b$ below the bifurcation value,
vacancy breathers do not exist.

\section{Small and large amplitude breathers in  Fermi-Pasta-Ulam lattices}

This section is dedicated to the study of breathers in
Fermi-Pasta-Ulam (FPU) lattices, they can be considered as a
subclass of the more generic Klein-Gordon lattices, in which the
on-site potential term has been set to zero. An FPU lattice can be
simply defined as a one-dimensional chain of particles subjected
to nonlinear interactions between them. Therefore its Hamiltonian
is given by
\begin{equation}
\label{eq:Hfpu} H= \sum_n \left[\frac{1}{2}
\dot{x}_n^2+W(x_n-x_{n-1}) \right],
\end{equation}
where $x_n$ describes the displacement of a particle from its
equilibrium position, and $W(x)$ is a nonlinear interaction
potential satisfying $W(0)=W'(0)=0$, $W''(0)>0$. A characteristic
of this system is that it does not possess the so called
"uncoupled limit", for which there exist trivial breathers.
  Although some early results on DBs  were obtained in FPU chains
in the late eighties by Sievers and Takeno ~\cite{Sievers}, a
complete and rigorous proof of their existence in these type of
lattices was provided only some years ago. In 2001, using a
variational method, Aubry {\it et al.} ~\cite{Aubry01} proved the
existence of breathers with frequency above the phonon spectrum if
the non linear interaction potential, $W$, is a strictly convex
polynomial of degree 4. In the same year and based on a center
manifold technique, James ~\cite{James01} proved the existence
(respectively, nonexistence) of small amplitude breathers (SAB)
with frequency slightly above the phonon band if $W$ satisfies
(respectively, violates) the following local hardening condition:
\begin{equation}
\label{defB} B=\frac{1}{2} W^{(4)}(0)-(W^{(3)}(0))^{2}>0.
\end{equation}
In spite of the implicit character of both proofs, the center
manifold method provides additional information on the breathers
amplitudes. In fact the method shows that for the interaction
force $y_n=W'(u_n)$, all small amplitude time-periodic solutions
with frequency $\wb$ of the FPU system (including breathers) have
the form
\begin{equation}
\label{solexact} y_{n}(t)= b_{n}\cos{(\omega_b\, t)}+
\mbox{h.o.t},
\end{equation}
being the amplitudes $b_n$ determined by the two-dimensional map
\begin{equation}
\label{map} b_{n+1}+2b_{n}+b_{n-1}= -\mu\,b_{n}+B\, b_{n}^{3}+
\mbox{h.o.t},
\end{equation}
provided $\wb$ lies near the upper edge, $\omega_{\pi}$, of the
phonon band, i.e. provided $\mu=\wb^2-\omega_{\pi}^2\ll 1$.  For
$B>0$ fixed and $\mu>0$ small enough, DBs correspond to homoclinic
solutions to 0 of the recurrence relation (\ref{map}) satisfying
$\lim\limits_{n\rightarrow\pm\infty}{b_n}=0$. In Refs
~\cite{James01} and ~\cite{James03} an existence proof is given
for two homoclinic solutions to 0, denoted as $\pm\, b^1_n$,
$\pm\, b^2_n$, having different symmetries $b^1_{-n-1}=-b^1_n$
(bond-centered mode) and $b^2_{-n}=b^2_n$ (site-centered mode).

  Since $0<\mu\ll 1$ Eq.~(\ref{map}) can be approximated at leading
order by an integrable differential equation ~\cite{Bernardo04}
and it is easy to find the following approximations to the exact
solutions $y_n^1$ and $y_n^2$:
\begin{eqnarray}
y^1_n(t)&\simeq& (-1)^n \sqrt{\frac{2\mu}{B}} \frac{\cos \wb
t}{\cosh(\, (|n+1/2|-1/2)\, \sqrt{\mu}\, )} , \label{eq:SAB} \\
y^2_n(t)&\simeq& (-1)^n \sqrt{\frac{2\mu}{B}} \frac{\cos \wb
t}{\cosh(n\sqrt{\mu})} \label{eq:SAB2}
\quad . \end{eqnarray} %
These expressions show  that the maximum amplitude  of SAB
is %
\begin{equation} A\approx \sqrt{\frac{2\mu}{B}} \label{eq:A}\end{equation}
while their width is $O(\mu^{-1/2})$ and diverges as $w_b
\rightarrow \omega_{\pi}^{+}$.

It is important to check the range of validity of the center
manifold approximations (\ref{eq:SAB}), (\ref{eq:SAB2}). A
detailed numerical study of DBs in an FPU chain has been done
~\cite{Bernardo04} with the anharmonic potential
\begin{equation}
W(u)=\frac{u^2}{2}+\frac{K_3}{3} u^3+\frac{K_4}{4} u^4 \quad
\label{eq:W}
\end{equation}
and fixing $K_4=1$. With this choice $B=3-4 K_3^2$, and thus the
parameter $B$ is positive if $|K_3|<K_3^{*}=\sqrt{3}/2\simeq
0.86$. In the case $K_3=0$ the potential is even with $B=3>0$. For
numerical convenience it is better to use as dynamical variables
the difference displacements $u_n=x_n-x_{n-1}$. With these
variables the equations of motions become
\begin{equation}
 \label{eq:dynu}
\ddot{u}_n+2W'(u_n)-\left[W'(u_{n+1})+W'(u_{n-1})\right]=0 \;
,\quad  n\in \mathbb{Z} \;.
\end{equation}
Note that $u_n$ is one-to-one related to the forces $y_n$ at small
amplitudes, because $W'$ is locally invertible since
$W^{\prime\prime}(0)\neq 0$. Indeed, expressions (\ref{eq:SAB}),
(\ref{eq:SAB2}) also approximates $u_n$  for small $\mu$ since
$u_n =y_n +O(y_n^2)$. Moreover ussing periodic boundary conditions
$u_{n+2p}(t)=u_{n}(t)$ the maximum frequency of the linear
phonons, $\omega_{\pi}$, is exactly 2 as in the infinite lattice.

\begin{figure}
    \begin{center}
            \includegraphics[clip=true,width=\singlefig]{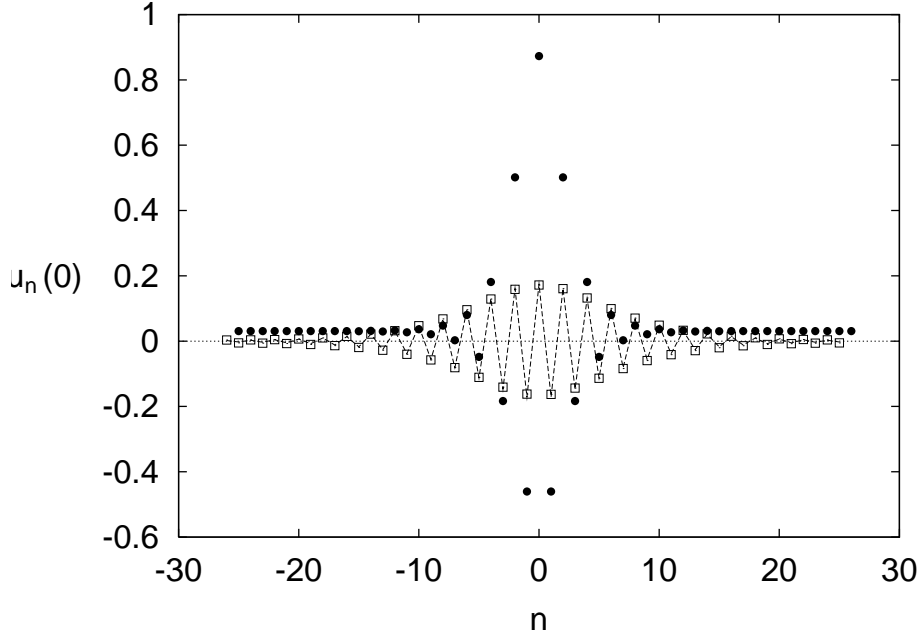}
    \end{center}
    \caption{Comparison between a small amplitude breather (squares) for $B>0$ and  a large
amplitude breather (full circles) for $B<0$ with the same
frequency $w_b=2.01$. This last breather is superposed to an
uniformly static state given by $\lim_{n\rightarrow \pm \infty}
u_n\approx 0.03$. The dashed line represents the centre manifold
approximation (\ref{eq:SAB2}) for small amplitude breathers.
Parameters: SAB: $K_3=-0.3$, $K_4=1$ ($B=2.64$); LAB: $K_3=-1$,
$K_4=1$ ($B=-1$).} \label{Fig11}
\end{figure}

    As can be appreciated in Fig.~\ref{Fig11}, there exist an excellent
 agreement between the approximation~(\ref{eq:SAB2}) (dashed line) and
 an exact site-centered mode obtained numerically (squares) for $\wb=2.01$
($\mu\approx 0.04$) and $K_3=-0.3$. Continuing this SAB as $\wb$
goes away from the phonon band, according to~(\ref{eq:A}), the
maximum amplitude of the relative displacements is approximately a
linear function of $\mu^{1/2}$ up to $\mu \approx 2$, relatively
far from the parameter region in which the center manifold
approximation is valid in principle. In fact, the results show
that, surprisingly, expressions (\ref{eq:SAB}) and (\ref{eq:SAB2})
fit very well the profile of the relative displacements $u_n$ even
for breathers with moderate amplitudes.


The linear stability of the SAB can be studied through a Floquet
analysis ~\cite{A97}.
 In the symmetric potential case the result is well known ~\cite{Bickham}:
the Page mode (site-centred mode in the $u_n$ variables) is stable
while the Sievers-Takeno mode (bond-centred mode in the $u_n$
variables) has a harmonic instability (a pair of real eigenvalues
$\sigma ,\sigma^{-1}$ close to $1$) that increases with the
breather frequency. When a cubic term in the potential $W$ is
introduced $(K_3\ne 0)$, the situation is more complex. Non-even
potentials induce oscillatory instabilities in both modes, which
increase with $|K_3|$. Although most of the oscillatory
instabilities are size dependent effects, some of them survive in
the limit of an infinite lattice. Thus in the non-even potential
case, both modes turn out to be unstable  ~\cite{Flach04}.
 The recurrence relation (\ref{map}) assures  that SAB with frequencies
slightly above the phonon band do not exist for $B<0$
($\sqrt{3K_4}/2<|K_3|$). However, in this parameter region
numerical simulations show that there exist breathers whose
amplitudes do not tend to zero as $w_b \rightarrow 2^{+}$
~\cite{Bernardo04}. As a consequence the energies of this family
of breathesrs lay always above a certain positive lower bound. The
existence of such an energy threshold is a rarely observed
phenomenon in one-dimensional lattices.~\cite{Kastner}

    The breathers obtained for $B<0$ have the same symmetries
(site-centred and bond-centred modes) and  the same stability
properties as SAB with non-even potentials, i.e. both modes show
oscillatory instabilities and also a harmonic instability in the
case of the Sievers-Takeno mode.

Another interesting problem is the study of breathers superposed
to uniformly stressed static states of the FPU chain given by
$x_n=c\, n$, being $c$  a nonzero constant ~\cite{Sandusky}.
Introducing the change of variables $x_n (t)=c n+\tilde{x}_n(t\,
\sqrt{{W}^{\prime\prime}(c)})$, the previous analytical and
numerical analysis can be completely applied to the renormalized
displacements $\tilde{x}_n$ ~\cite{Bernardo04}. As a consequence,
it is easy to prove that the FPU model has a family of exact
breather solutions satisfying $\lim_{n\rightarrow \pm \infty}
u_n=c$, which can be approximated by
\begin{equation}
u_n(t)\simeq c+(-1)^n \sqrt{\frac{2\mu}{B(c)}} \frac{\cos (\wb t
)}{\cosh(n\sqrt{\mu  /{W}^{\prime\prime}(c) })} \label{eq:SABc},
\end{equation}%
where $\mu =\omega_b^2-4\, {W}^{\prime\prime}(c) \ll 1$, and
$B(c)=\frac{1}{2} {W}^{\prime\prime}(c)\,
W^{(4)}(c)-(W^{(3)}(c))^{2}>0$. An example of a breather with a
uniform stress is shown in Fig.~\ref{Fig11} with full circles.
Note that the lower frequency of these breather solutions lies
inside the phonon band if ${W}^{\prime\prime}(c) <1$, and above it
for ${W}^{\prime\prime}(c)>1$.

\section{Dark breathers in Klein--Gordon lattices}

The study of breathers in chains of nonlineraly interacting
particles has a natural extension in the study of other localized
entities called "dark breathers". The term dark breather refers to
a state of the chain where most of the oscillators are excited
except one or a few units of them which have a very small
amplitude. That is, in some sense it posses the opposite
properties than a breather. This section summarizes some results
about the conditions for the existence and stability properties of
dark breathers in different Klein--Gordon lattices.

 Consider one--dimensional Klein--Gordon lattices of oscillators with Hamiltonian
\begin{equation}
  H=\sum_{n} (\frac{1}{2}\,\dot{u}_n^2+V(u_n))+ \varepsilon\,
  W(u),
\label{eq:ham1}
\end{equation}
 where $u_n$ represents the coordinates of the oscillators
referred to their equilibrium positions; $V(u_n)$ represents the
on--site potential; $u$ represents the set of variables $\{u_n\}$;
and $\varepsilon W(u)$ represents the interacting potential.
 The parameter $\varepsilon$ describes the intensity of the coupling,
 and $ W(u)$ is given by

\begin{equation}
W(u)=\frac{1}{2}\,\sum_{n}(u_{n+1}-u_n)^2.
\end{equation}
This interaction is attractive for $\epsilon>0$, as a nonzero
value of a variable tends to increase the values of the
neighbouring variables with the same sign. The on--site potential
is given by
\begin{equation}
V(u_n)=\frac{1}{2}\,\omega_0^2\;u_n^2+\,\phi(u_n),
\end{equation}
with $\phi(u_n)$ being the anharmonic part of the potential. The
variables can be scaled so that all the particles in the lattice
have mass unity and the linear frequency $\omega_0=1$.

The existence theorem by Mackay and Aubry ~\cite{MA94} establishes
that dark breather solutions are possible, at least up to a
certain value of the coupling parameter $\varepsilon_c$. Dark
breathers can be calculated numerically. The method to calculate
this type of solutions is similar to the used for obtaining
breathers ~\cite{AACR02}. Fig.~ref{Fig12}, shows two different
examples of dark breather profiles for a chain with a cubic soft
on--site potential, with attractive interaction (left panel), and
with repulsive interaction (right panel).


\begin{figure}
  \begin{center}
\includegraphics[clip=true,width=\singlefig]{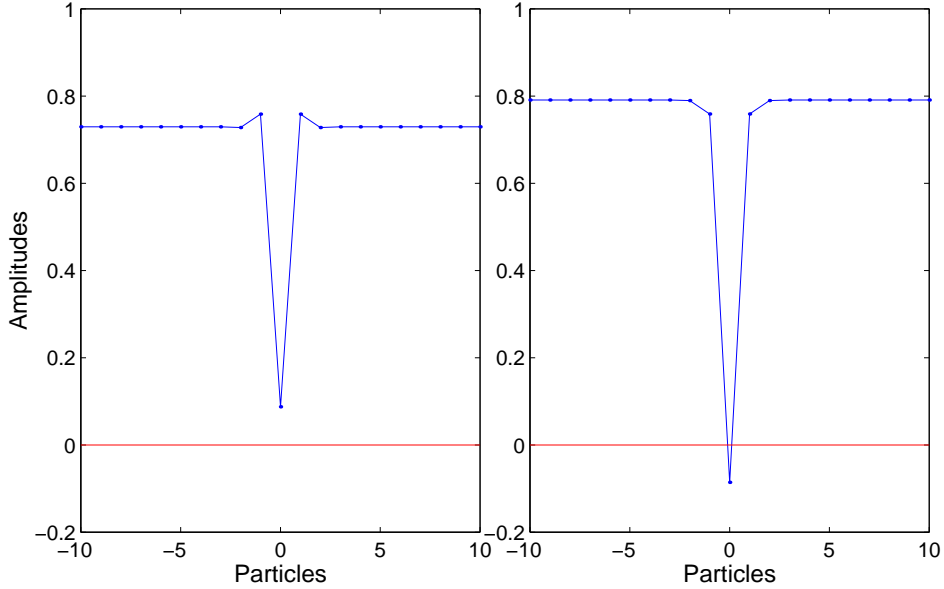}
   \end{center}
  \caption{Dark breathers profile for a cubic potential with
  attractive interaction (left) and with repulsive
  interaction(right), with $\varepsilon=0.023$.}
  \label{Fig12}
\end{figure}


The linear stability analysis of dark breather solutions can be
performed numerically. For that porpose both the Floquet
multipliers and the Aubry band theory can be considered
~\cite{A97,MAF98}.

\begin{figure}
  \begin{center}

    \includegraphics[clip=true,width=\singlefig]{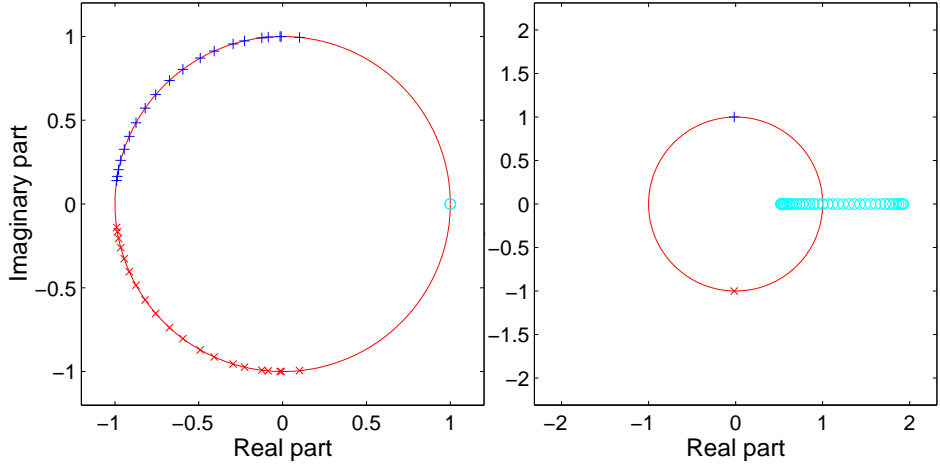}
  \end{center}
  \caption{Evolution of the Floquet multipliers with cubic
  on--site potential and attractive interaction for: (left) a
  breather with $\varepsilon=0.1$; (right) a dark breather with
  $\varepsilon=0.004$. The breather frequency is in both cases $\wb=0.8$.}
  \label{Fig13}
\end{figure}

In the case of a soft on--site potential of the type
$\phi(u_{n})=-\frac{1}{3}u_{n}^{3}$ , and attractive interaction
between particles, dark breather solutions are not stable.  Fig.
\ref{Fig13} shows the type of stability of a dark breather and a
breather by the Floquet multipliers method. The breather (left) is
stable for $\epsilon\leq0.1$, whereas the dark breather (right) is
unstable for all the $\epsilon$ values considered. This
instability is due to harmonic bifurcations.


\begin{figure}[h]
  \begin{center}
    \includegraphics[clip=true,width=\singlefig]{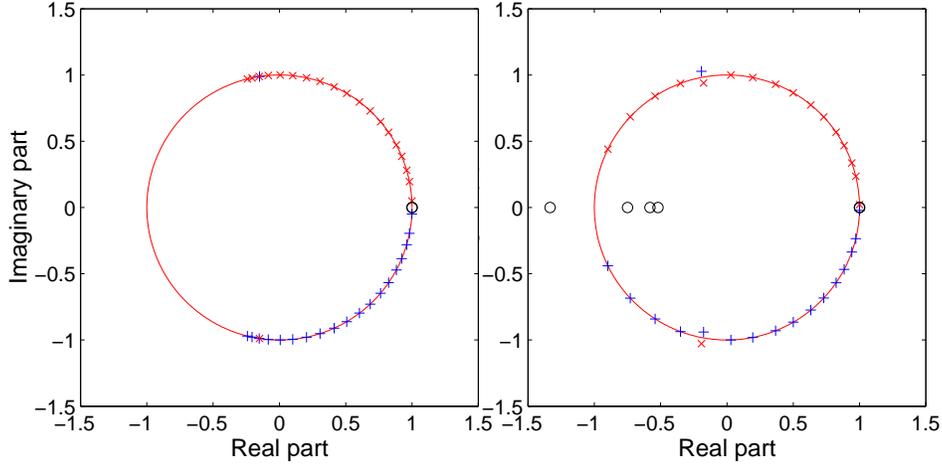}
   \end{center}
  \caption{Floquet eigenvalues for a Morse
  on--site potential and repulsive coupling. Left: $\varepsilon=0.024$,
  the system is still stable. Right: $\varepsilon=0.035$,
  the system becomes unstable due to oscillatory and subharmonic bifurcations.}
  \label{Fig14}
\end{figure}


 Considering chains with repulsive interactions, the linear stability analysis
shows that dark breather solutions are stable up to a significant
value of the parameter $\epsilon$.

The type of stability behaviour described above appears also in
other kind of soft on--site potentials as, for example, the Morse
potential given by
\begin{equation}
V(u_n)=D\,(\exp(-bu_n)-1)^2.
\end{equation}
 Dark breather solutions are stable if the coupling
is repulsive for $\epsilon\leq0.024$. For bigger values the
solution becomes unstable due to oscillatory and subharmonic
bifurcations ~\cite{AACR02}. Fig.~\ref{Fig14} shows the evolution
of the Floquet multipliers for $\epsilon=0.024$ (left panel),
where the dark breather solution is still stable, and for
$\epsilon>0.024$ (right panel), where oscillatory and subharmonic
bifurcations appear.



\begin{figure}[t]
  \begin{center}
    \includegraphics[clip=true,width=\singlefig]{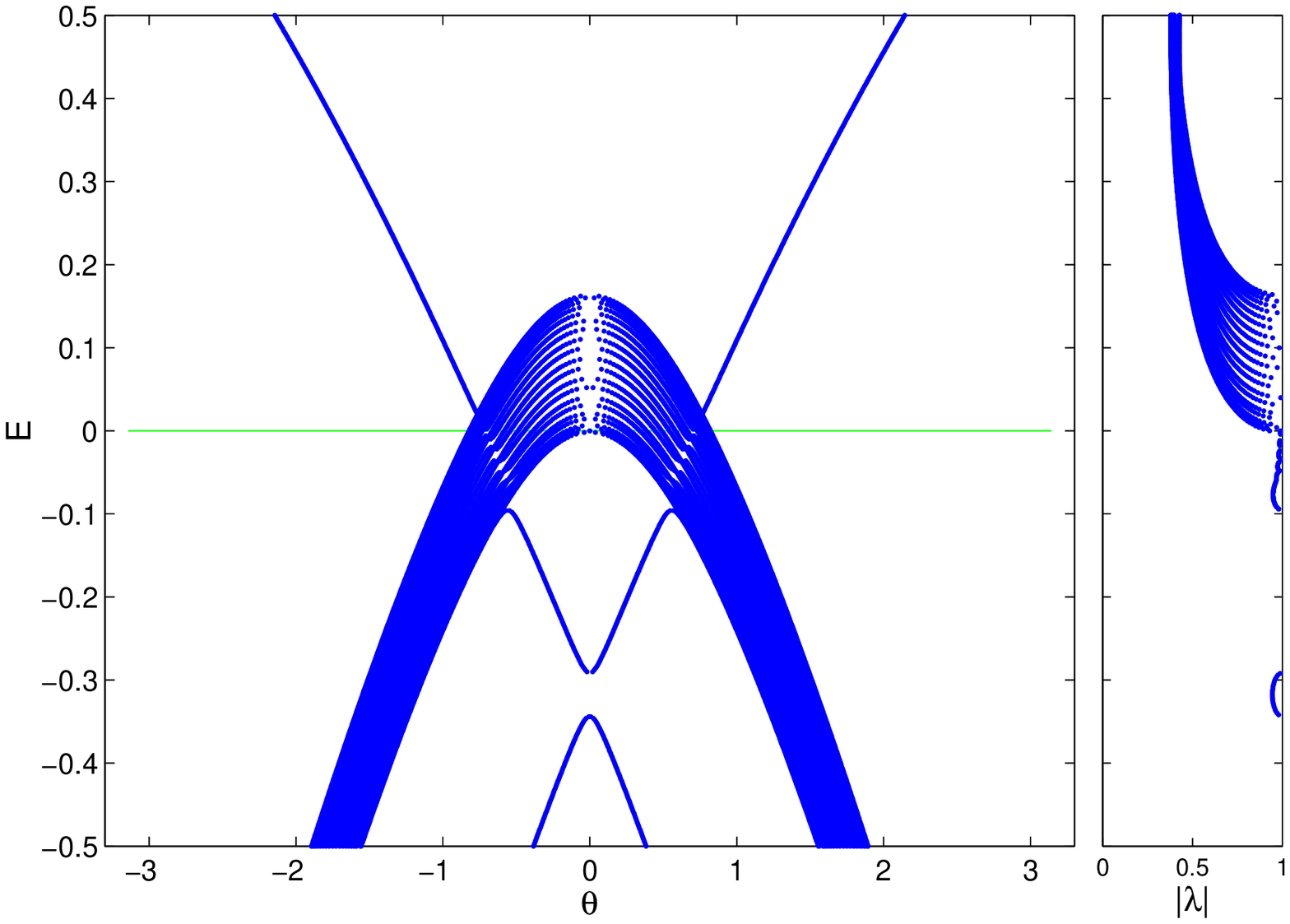}
  \end{center}
  \caption{Band structure for a quartic hard on--site potential
   with attractive interaction at $\varepsilon=0.041$ and $\wb=1.2$.}
  \label{Fig15}
\end{figure}

 Lattices with hard on--site potential present a different scenario.
The dark breather solution is stable with attractive coupling
potential, up to a certain value of $\epsilon$, and is unstable if
the coupling is repulsive. Oscillatory, subharmonic and harmonic
bifurcations appear depending on the form of the on--site
potential. For a hard on--site potential of the type
$\phi(u_{n})=\frac{1}{4}u_{n}^{4}$ and attractive coupling, the
system is stable for $\epsilon\leq0.022$. Figs. ~\ref{Fig15} and
~\ref{Fig14} represent, respectively, the band structure and the
Floquet multipliers at $\epsilon=0.041$. They show that the dark
breather solution is unstable due to an harmonic and small
oscillatory bifurcations. The instability mode, shown in
Fig.~\ref{Fig15} (right panel), is an asymmetric extended one.
Simulations performed perturbing with it the dark breather give
rise to a small oscillation with both sides of the chain out of
phase, superimposed on the dark breather one, but the darkness is
preserved. After that, there are oscillatory bifurcations due to
the band mixing with the usual properties.

An analysis of larger systems shows that subharmonic and
oscillatory bifurcations persist even though the system is
infinite, while in the case of breathers there are two kinds of
size--dependent bifurcation ~\cite{MA98}.
\begin{figure}[t]
  \begin{center}
    \includegraphics[clip=true,width=\singlefig]{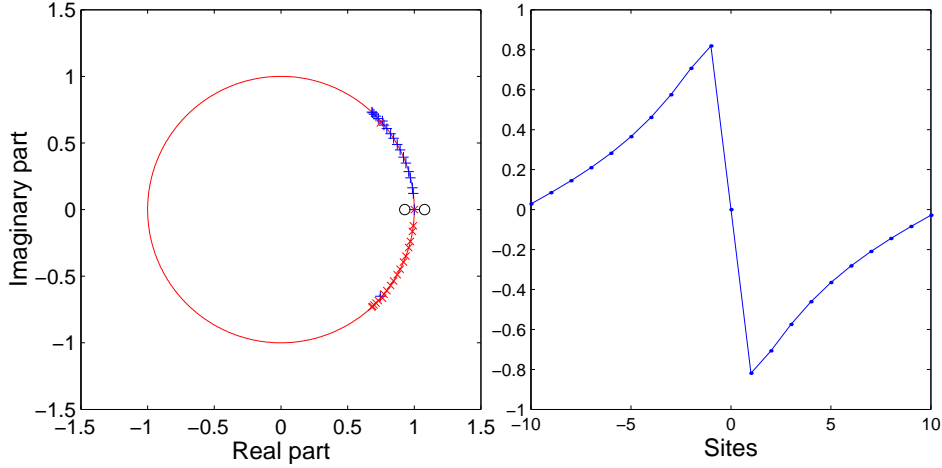}
  \end{center}
  \caption{Left: Floquet multipliers with a quartic hard on--site potential
   and attractive interaction at $\varepsilon=0.041$ and $\wb=1.2$. A
   harmonic and a small oscillatory instabilities appear. Right: the
   velocities components (the position ones are zero) of the eigenvalue
    corresponding to the harmonic instability.}
  \label{Fig16}
\end{figure}

\section{Polarons in biomolecules}

 The study of charge transport (CT) in biomolecules, and particularly in
DNA, is of utmost importance. Electronic transport plays a very
important role in many biological processes, such as DNA repair
after damage by radiation or chemical agents, photosynthesis,
nevous conduction, etc. Also, because the quasi--one--dimensional
structure of DNA and other biomolecules could be used to design
electronic devices based on biomaterials (for a detailed
bibliographical information, see i.e. ~\cite{Hen04}).

 When an electron moves along the double
strand of DNA it is accompanied by a local deformation of the
molecule, forming a polaron or an electron--vibron breather. The
polaronic character of the CT along DNA has been studied by means
of different models, some of then based on the Peyrard--Bishop DNA
model ~\cite{PB89}, in which the only variables are the distances
between bases within each base pair. have permitted to show that
the polaronic charge transport along DNA can be mediated by the
coupling of the charge carrying unit with the hydrogen--bond
deformations,

In particular, it has been shown in some simple, but powerful
models based on the  proposed by Peyrard--Bishop ~\cite{PB89},
that the polaronic CRT along DNA can be mediated by coupling of
the charge carrying unit with the hydrogen--bond deformations, and
that this transport survives to a small amount of parametric and
structural disorder ~\cite{HAA03}, or by means of the coupling of
this charge unit with the twist deformation of neighbor base pairs
~\cite{PAHR03}. In this last case, the moving polaron is slower
than in the previous one, but it is more robust under the
introduction of parametric disorder. These differents regimes
apperars for different values of a parameter $\alpha$, the couples
the transference integral with the deformations of the hydrogen
bonds, a parameter which at present there is not any realible
experimental data.

 Consider a three-dimensional semi--classical tight--binding
model for synthetically produced DNA polymers built up from single
type of base pairs, i.e. either poly(dG)-poly(dC) or
poly(dA)-poly(dT) DNA polymers ~\cite{Porath}. Utilizing a
nonlinear approach based on the concept of breather and polaron
solutions it is possible to explore if the conductivity depends on
the type of the DNA polymer. If they have quantitatively distinct
transpor properties or not might be of interest for the design of
synthetic molecular wires. Moreover, it has been possible to
ameliorate preceding studies of electronic transfer in DNA
~\cite{HAA03},~\cite{control} in the sense that, instead of
adjusting the coupling parameters, plausible estimates for them,
derived with the help of quantum-chemical methods, has been used.

\begin{figure}[t]
\begin{center}
 \includegraphics[clip=true,width=\singlefig]{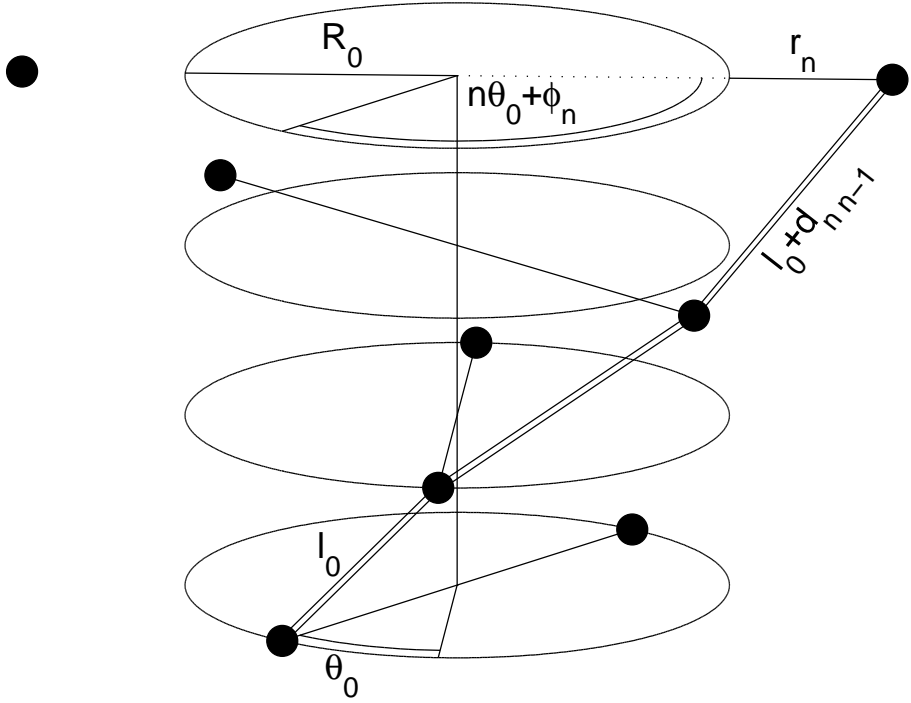}
  \caption{Sketch of the structure of the DNA model. The bases are
  represented by bullets.}
  \label{Fig17}
\end{center}
\end{figure}

 The model for CT in DNA is based on the finding that
the charge migration process is dominantly influenced by the
transverse vibrations of the bases relative to each other in
radial direction within a base pair plane. The impact of other
vibrational degrees of freedom are expected to be negligible with
respect to charge transport in DNA. Then, the motion can be viewed
as confined to the base pair planes ~\cite{Stryer}. The
Hamiltonian for CT in along a strand in DNA has two parts
$H=H_{el}+H_{vib}$, with $H_{el}$ describing the CT over the base
pairs and $H_{vib}$ the dynamics of radial vibrations of the base
pairs or the vibronic part. A sketch of the structure of the DNA
model is shown in Fig.~\ref{Fig17}.


Supposing that vibrational dynamics can be described classically,
it is easy to formulate the dynamical equations of the system (a
detailed description of procedure can be found in
~\cite{PAHR03}).It is possible to use some typical parameter
values for DNA molecules and compute plausible values for the
electron--coupling strengths as a result of quantum--chemical
computational procedure ~\cite{Hen04}.

The nonlinear interplay between the electronic and the vibrational
degrees of freedom, makes possible the formation of polaronic
electron--vibration compounds. Also, there exist some stationary
solutions that can be calculated with the help of the nonlinear
map approach explained in detail in ~\cite{Kalosakas}. In general,
the polarons are of fairly large extension (width). Regardless of
the DNA polymer type, the electronic wave function is localized at
a lattice site and the envelope of the amplitudes decays
monotonically and exponentially as distance grows from this
central site (base pair). However, the electronic wave function of
the poly(dG)-poly(dC) DNA polymer is stronger localized than the
one of its poly(dA)-poly(dT) counterpart. As polarons solutions
are fairly large extended (the half-width involves $\lesssim 100$
lattice sites) it is plausible to think that the can be mobile.
This polaron motion can be activated using the discrete gradient
method to obtain suitable initial perturbations of the momentum
coordinates which initiate coherent motion of the polaron
compound.
 The electronic occupation probability is defined as
$\bar{n}(t)=\sum_n\,n\,|c_n(t)|^2$, where the index $n$ denotes
the site of the $n-$th base on a strand and $c_n$ determines the
probability to find the electron (charge) residing at this site.

\begin{figure}[h]
\begin{center}
 \includegraphics[clip=true,angle=-90,width=0.7\textwidth]{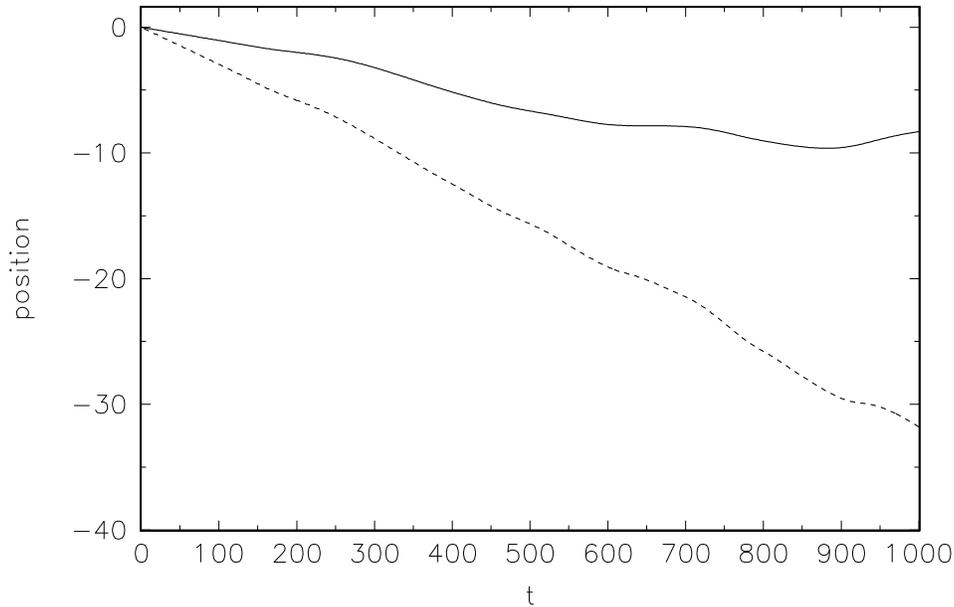}
  \caption{ Time evolution of the first momentum of
the electronic occupation probability. Full (dashed) line:
poly(dA)-poly(dT) (poly(dG)-poly(dC)) DNA polymer.}
  \label{Fig18}
\end{center}
\end{figure}

  Results relative to the time evolution of the first momentum of the
electronic occupation probability are shown in Fig.~\ref{Fig18}.
Conductivity in synthetically produced DNA molecules depends on
the type of the single base pair of which the polymer is built of.
While a polaron-like mechanism, relying on the nonlinear coupling
between the electron amplitude and radial vibrations of the base
pairs, is responsible for long-range and stable electronic
transfer in (dG)-(dC) DNA polymers, the conductivity is
comparatively weaker in the case of (dA)-(dT) DNA polymers.
Especially when it comes to designing synthetic molecular wires
these findings might be of interest. In fact, recent experiments
suggest that electronic transfer through DNA molecules proceeds by
polaron hopping ~\cite{Lee}. Furthermore, these results comply
with the findings of experiments which show also that
poly(dG)-poly(dC) DNA polymers forms a better conductor than their
poly(dA)-poly(dT) counterparts.


\section{Quantum breathers}

At present, the phenomenon of localization of energy due to
nonlinearity in discrete classical lattices is relatively well
understood, but the understanding of the quantum equivalence of
discrete breathers is very limited. There exist some theoretical
results ~\cite{seg94,QTrev} and some  experimental observations of
these states in different quantum systems, as mixtures of
4--methyl--pyridine ~\cite{Fil}, in Cu benzoate ~\cite{Asa00}, and
in doped alkalihalides ~\cite{Sch02}. Also, the study of discrete
breathers in quantum lattices supporting a small number of quanta
is not only interesting theoretically, this knowledge may be
relevant to studies of quantum dots and quantum computing
~\cite{li03}, and for studies of Bose--Einstein condensates
trapped in optical traps ~\cite{BEC}.

 This section focus the attention in one--dimensional lattices with a
small number of quanta described by the the quantum version of the
discrete nonlinear Schr\"{o}dinger equation (QDNLS), also known as
the Hubbard model. This is a particularly simple model for a
lattice of coupled anharmonic oscillators, which has been used to
describe the dynamics of a great variety of systems ~\cite{sc99}.
We present the results corresponding to a periodic lattice with
$f$ sites containing bosons ~\cite{eil04}. Many of these results
can be extended to a great variety of systems, for example,
similar results have been obtained with a periodic lattice
containing fermions, described by an attractive fermionic Hubbard
model (FH) with two kinds of particles with opposite spins. This
is a model of interest in connection with the theory of high-$T_c$
superconductivity, and it can be used to describe bound states of
electron and holes in some nanostructures as nanorings (excitons)
~\cite{eil03}.

The QDNLS Hamiltonian of the quantum system is
\begin{equation}
\hat H=-\sum_{j=1}^{f} \frac12\gamma_j b_j^\dag b_j^\dag b_j b_j+
\epsilon_j b_j^\dag(b_{j-1}+b_{j+1}), \label{Ham_bos}
\end{equation}
where $b_j^\dag$ and $b_j$ are standard bosonic operators,
$\gamma_j/\epsilon_j$ is the ratio of anharmonicity to nearest
neighbor hopping energy, and periodic boundary conditions are
imposed. As first step, only short range interactions are
considered given by the hopping coefficients $\epsilon_j$.

This Hamiltonian conmutes with the number operator
$N=\sum_{j=1}^{f} b_j^\dag b_j$, and can be block-diagonalised
easily when the number of bosons is small enough. Numerically
exact solutions can be found restricting the study to small
lattices with a small number of quanta.  The simplest nontrivial
case is for $N=2$, where bound states corresponds to bound
two-vibron states, as has been observed experimentally in several
systems ~\cite{Jak02}. These results have been extended to more
complicated situations, noting that many of them are valid for
larger values of $N$.

\subsection{Quantum breathers in a translational invariant lattice}

In a homogeneous quantum lattice the parameters $\gamma_j$ and
$\epsilon_j$ are independent of $j$, i.e. $\gamma_j=\gamma$ and
$\epsilon_j=\epsilon$. With periodic boundary conditions, it is
possible to block--diagonalize the Hamiltonian operator using
eigenfunctions of the translation operator $\hat T$, defined as
$\hat T b_j^\dag = b_{j+1}^\dag \hat T$. In each block, the
eigenfunctions have a fixed value of the momentum $k$, with
$\tau=\exp(i k)$ being an eigenvalue of the translation operator
~\cite{sc99}. In this way, it is possible to calculate the
dispersion relation $E(k)$ with a minimal computational effort.

\begin{figure}[h]
\begin{center}
 \includegraphics[clip=true,width=\singlefig]{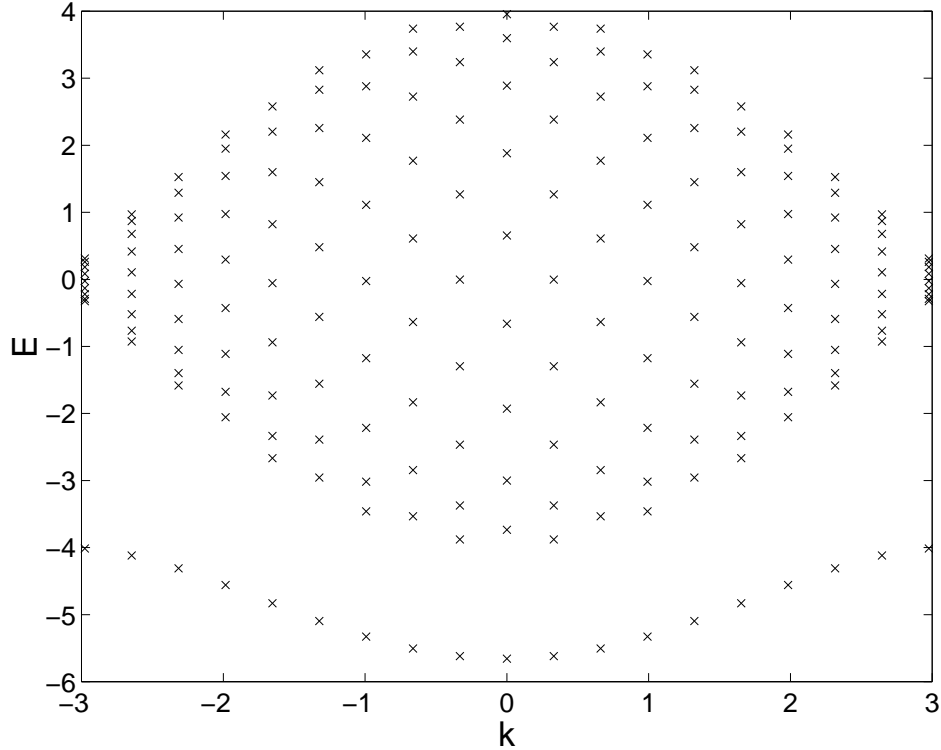}
  \caption{Eigenvalues $E(k)$. $N=2$, $f=19$ and $\gamma=4$.}
  \label{Fig19}
\end{center}
\end{figure}

In general, as shown in ~\ref{Fig19}, if parameter $\gamma$ is
high enough, there exists an isolated ground state eigenvalue for
each $k$ which corresponds to a quantum breather.  In this state,
there is a high probability of finding the quanta on the same
site, but due to the translational invariance of the system, there
exists an equal probability of finding these quanta at any site of
the system. In these cases, some analytical expressions can be
obtained in some asymptotic limits ~\cite{sc99}.

For example, using the number--state--basis $
|\psi_n>=[n^a_1,n^a_2,...,n^a_f]$, where $n^a_i$ represents the
number of quanta at site $i$, in the case $N=2$ and $k=0$ the
(unnormalized) ground state is given by $$
|\Psi\rangle=[20\dots0]+[020\dots0]+\dots+[0\dots02]+O(\gamma^{-1}).
$$ The coefficients $a_i$ of the first $f$ terms are equal to
unity and the rest are $O( \gamma^{-1})$.

\subsection{Trapping in lattices with broken translational
symmetry}

 When the translational invariance of a lattice is broken, the Hamiltonian
operator does not commute with the translation operator and
breathers can be localized around a particular point of the
lattice. Perhaps the simplest way to do it is to consider
non--flux boundary conditions, in order to simulate a finete--size
chain. In this case, the ground state becomes localized around the
middle of the chain but, when $f$ is high enough, this effect
vanishes.

The existence of local inhomogeneities or impurities can also
break the translational symmetry of the system, and this can be
modeled by making some coefficients dependent on the site. This
can occurs in system with twisted or bent geometries, as  in
models of globular proteins ~\cite{ei86} or photonic crystals
In these cases, the geometry effects can be considered
introducing a long--range hopping term in the Hamiltonian
operator.

\begin{figure}[h]
\begin{center}
 \includegraphics[clip=true,width=\singlefig]{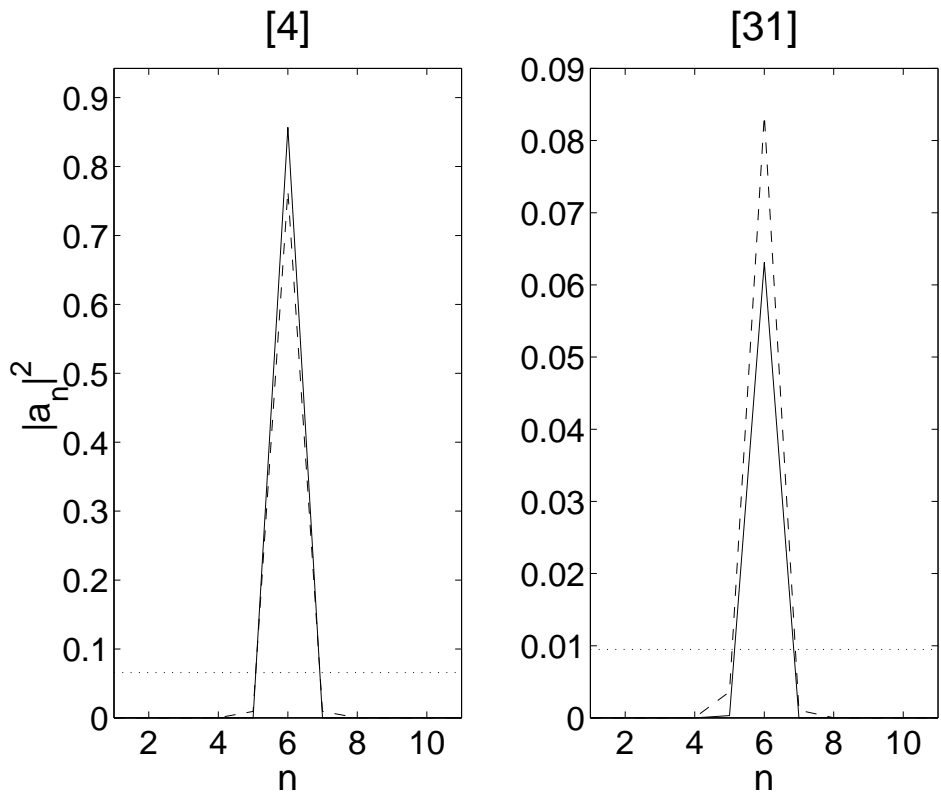}
  \caption{Square wave function amplitudes corresponding to
    localized components of the ground state. $f=11$, $N=4$ and $\gamma=2$.
    Point impurity at the site $\ell=6$. ($\cdot\cdot\cdot$)
    homogeneous chain case, $\gamma_\text{im}=2$, (- - -)
    $\gamma_\text{im}=2.1$ , (---) $\gamma_\text{im}=2.5$.}
  \label{Fig20}
\end{center}
\end{figure}

 The breaking of translational symmetry of the system localizes
the breather state around a particular site of the chain, as shown
in Fig. ~\ref{Fig20}. Note that the localization phenomena due to
random variation of lattice parameters have been studied in
harmonic models since the work of Anderson ~\cite{an58} but, in
all cases, and due to anharmonic effects, there exist new
localization effects, and, when it is present, the anharmonicity
enhances the Anderson--like effect.

\section{ Thermal evolution of enzyme-created oscillating bubbles in DNA}

This last section reviews some fundamental researches on the
creation and subsequent evolution of breathers in DNA chains. For
a double--stranded DNA molecule to replicate, the two strands of
the double helix must be separated from each other, at least
locally, with the breaking up of the hydrogen bonds. For long
double--stranded DNA molecules, the rate of spontaneous strand
separation is very low under physiological condition. Specific
enzimes belonging to the large family  of nucleic acid helicases
(see ~\cite{Stryer}) (there are more than sixty different types),
utilize the energy of ATP hydrolysis to power strand separation.
The opening of a selected region of the twisted (Watson-Crick) DNA
double helix, begins with a partial unwinding at an area called
the replication fork and is observed as a bubble. Helicases are
ubiquitous and versatile enzymes that, in conjunction with other
components of the macromolecular machines, carry out important
biological processes such as DNA replication; DNA recombination;
nucleotide excision repair (i.e. UV damage); RNA editing and
splicing; the transfer of a single-stranded nucleic acid to other
nucleic acid or a protein or its release into solution
 ~\cite{Hippel01} and formation of T-loop telomeres~\cite{Wu01}.
 Nucleic acid helicases have several structural
varieties (monomeric, dimeric, trimeric, tetrameric and closed
hexameric), but all of them use the hydrolysis of nucleoside
triphosphate (NTP) to nucleoside diphosphate (NDP) as the
preferred source of energy. Helicases have also a huge interest
from a point of view of clinical disease. Actually, there are
known up to five human monogenic hereditary disorders in which the
mutated protein responsible for the disease carries an aberrant
helicase activity: cockayne 's syndrome, xeroderma pigmentosum,
trichothiodystrophy, Bloom 's syndrome, and Werner 's syndrome.
 The study of the formation of DNA openings has been initiated
recently utilizing models based on nonlinear lattice dynamics
 ~\cite{Englander,Agarwal}.

 With a view to vibrational excitations in DNA, the Peyrard--Bishop (PB)
 model ~\cite{PB89} and its successors ~\cite{Dauxois, BCP}
 have been successfully applied to describe moving localized excitations
(moving breathers), which reproduce typical features of the DNA
opening dynamics such as the magnitude of the amplitudes and the
time scale of the "oscillating bubble" preceding full strand
separation.

  The bubble formation process is initiated by structural
deformations of selected regions of the parental DNA duplex that
serves as a template for the replication. The process begins when
the replication apparatus identifies the starting point and then
gets combined to it. The replication apparatus is and assembled
complex that forms the replication fork and opens it
directionally. During this process the helical DNA is unwound and
replicated. One of the components of that complex a type of
helicase, which is unable to recognize the origin of the
replication {\it per se}, and which requires the participation of
specific proteins to lead it to the initiation site
~\cite{Delagoutte03}.

It is assumed that this enzyme operates in the way, that a local
unwinding in combination with (rather small) stretchings of the
H-bonds in this region occurs. A working model should demonstrate
that there are indeed initial deformations that give rise to
localized vibrations, constrained to a region of DNA, matching the
properties of oscillating bubbles observed experimentally. These
oscillating bubbles with their temporarily extended but yet
unbroken H-bonds serve as the precursors to the replication
bubble. Furthermore, it is important to know whether the stable
radial and torsional breathers persist under imposed thermal
perturbations.

The research has been done using a nonlinear oscillator network
model for the DNA double helix that is explained in detail in
~\cite{BCP,Cocco99,Agarwal}. The equilibrium position of each base
within the duplex configuration is described in a Cartesian
coordinate system by $x_{n,i}^{(0)}$, $y_{n,i}^{(0)}$ and
$z_{n,i}^{(0)}$. The index pair $(n,i)$ labels the $n$-th base on
the $i-$th strand with $i=1,2$ and  $1\leqq n\leqq N$\,, where $N$
is the number of base pairs considered. Displacements of the bases
from their equilibrium positions are denoted by $x_{n,i}$,
$y_{n,i}$ and $z_{n,i}$. The potential energy taking into account
the interactions between the bases consists of four parts. The
potential energy of the hydrogen bond within a base pair is
modeled typically by a Morse potential
\begin{equation}
V_{h}^n=D_n \left[\,\exp\left(-\frac{\alpha}{2}\,
d_n\right)\,-1\,\right]^2\,, \label{eq:Uhyd}
\end{equation}
where the variables $d_n$ describe dynamical deviations of the
hydrogen bonds from their equilibrium lengths $d_0$. The
site-dependent depth of the Morse potential, $D_n$, depends on the
number of involved hydrogen bonds for the two different pairings
in DNA, namely the G-C and the A-T pairs. The former pair includes
three hydrogen bonds while the latter includes only two.
$\alpha^{-1}$ is a measure of the potential--well width.

The energies of the rather strong and rigid covalent bonds between
the nucleotides $n$ and $n-1$ on the $i-th$ strand are modeled by
 harmonic potential terms
\begin{equation}
V_{c}^{n,i}=\frac{K}{2}\,l_{n,i}^2\,, \label{eq:Uback}
\end{equation}
and $l_{n,i}$ describes the deviations from the  equilibrium
distance between two adjacent bases on the same strand. $K$ is the
elasticity coefficient.

Effects of stacking, which impede that, due to the backbone
rigidity, one base slides over another ~\cite{Stryer} are
incorporated in the following potential terms
\begin{equation}
V_{s}^{n,i}=\frac{S}{2}\,(d_{n,i}-d_{n-1,i})^2\,.
\end{equation}
The supposedly small deformations in longitudinal direction can be
modeled by harmonic elasticity potential terms given by
\begin{equation}
V_{l}^{n,i}=\frac{C}{2}\,(z_{n,i}-z_{n-1,i})^2\,.
\end{equation}
The kinetic energy of a nucleotide is determined by
\begin{equation}
E_{kin}^{n,i}=\frac{1}{2m}\,\left[\,\left(p_{n,i}^{(x)}\right)^2
+\left(p_{n,i}^{(y)}\right)^2
+\left(p_{n,i}^{(z)}\right)^2\right]\,,\label{eq:Ekin}
\end{equation}
where $m$ is the mass and $p_{n,i}^{(x,y,z)}$ denotes the
$(x,y,z)-$component of the momentum.

The model Hamiltonian reads then as
\begin{equation}
H=\sum_{i=1,2}\,\sum_{n=1}^{N}\,E^{n,i}\,,\label{eq:Hdna}
\end{equation}
with
\begin{equation}
E_{n,i}=E_{kin}^{n,i}+V_{h}^{n}+V_{c}^n+V_{s}^{n,i}+V_{l}^{n,i}\,,\label{eq:Eloc}
\end{equation}
and the summation in (\ref{eq:Hdna}) is performed over all
nucleotides and the two strands.

\begin{figure}[t]
\centering
\includegraphics[clip=true,angle=270,width=\singlefig]{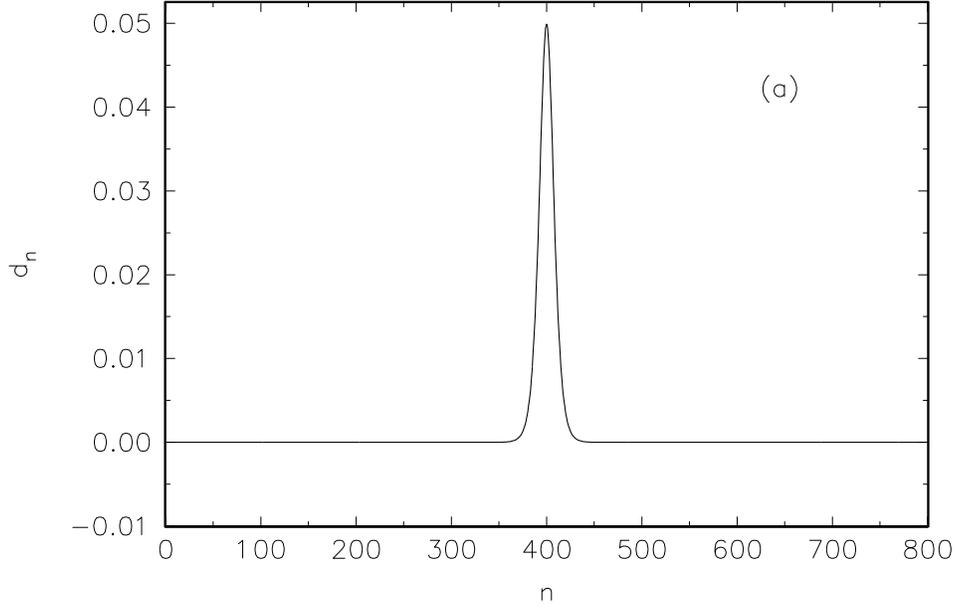}
\caption{ The initial radial distortions of the double helix. The
inter-base distance $d_n(0)$ in \AA.}
\label{Fig21}
\end{figure}

\begin{figure}[t]
\centering
\includegraphics[clip=true,angle=270,width=\singlefig]{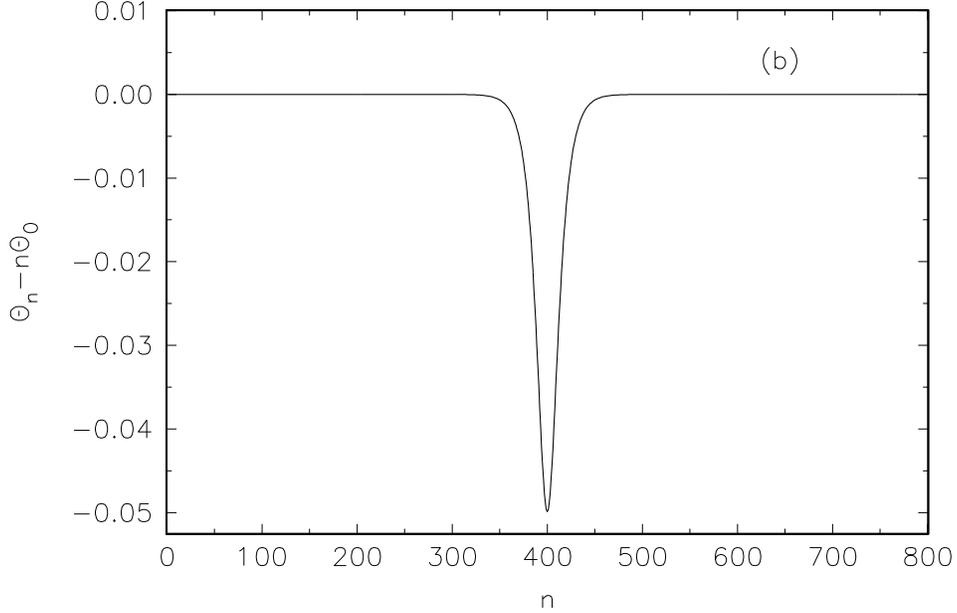}
\caption{The initial angular distortions of the double helix.}
 \label{Fig22}
\end{figure}

 Consider the initiation of the oscillating  bubbles in DNA. The
starting point is a DNA molecule for which a certain segment
experiences initially angular and radial deformations due to the
action of some helicase to which a region of the DNA is bound. In
order to simulate the deforming action of enzymes, assume that
initially a number of consecutive sites in the center of the DNA
lattice (hereafter referred to as the {\it central region}) are
exerted to forces acting in angular and radial direction such that
in this region the molecule experiences twist reduction together
with radial stretchings. These structural deformations can be
extended over a region encompassing up to thirty base pairs and as
can be demonstrated give rise to the formation of H-bridge
breather solutions (extending over $15-20$ base pairs) reproducing
the "oscillating bubbles" observed for the DNA-opening process
~\cite{Barbithesis}. In Figs.~\ref{Fig21} and \ref{Fig22} the
localized initial distortions are shown. Both the angular and
radial deformation patterns are bell-shaped, due to the fact that
the enzymatic force is exerted locally, i.e. in the extreme case
to a single base pair only for which the H-bond is deformed


The first one being of non-positive amplitudes is linked with
reduced twist while the latter one with non-negative amplitudes is
associated with radial stretchings. The radial and angular
deformation patterns are centered at the central lattice site
(base pair) at which the H-bond stretching and twist angle
reduction is at maximum. At either side of the central site the
amplitudes approach progressively zero. The deformation energy
amounts to $0.0362\,\mathrm{eV}$.
 The associated set of coupled equations has been
integrated with the help of a fourth-order Runge-Kutta method. For
the simulation the DNA lattice consists of $799$ sites and open
boundary conditions are imposed. The same initial conditions are
used both at zero and nonzero temperatures.

 The study of the temperature dependence of these oscillating bubbles
 has been done by means
of bringing in contact the nonlinear oscillator network with a
heat bath using the Nosé-Hoover-method. It is demonstrated that
the radial and torsional breathers sustain the impact of thermal
perturbations, even at temperatures as high as room temperature.
For T>0 the breathers start to travel coherently towards an end of
te DNA duplex, representing realistically the oscillating bubbles
found in NAs i.e. they oscillate with periods in the range 0.3
-0.8 ps, are on the average extended over 10 -20 base pairs and
possess maximal amplitudes of the order of 0.6 °A.

Quantitatively, the results regarding the temperature dependence
of the breather evolution are suitably summarized by the
time-evolution of the first momentum of the energy distribution
defined as
 \begin{equation}
\bar{n}(t)=\sum_{i=1,2}\,\sum_{n=1}^{N}\,(n_c-n)\,E_{n,i}(t)
\label{eq:momentum}\,,
\end{equation}
and the  energy $E_{n,i}$ is defined in Eq.\,(\ref{eq:Eloc}) and
$n_c$ is the site index corresponding to the center of the DNA
lattice.

\begin{figure}[t]
\centering
\includegraphics[clip=true,angle=270,width=\singlefig]{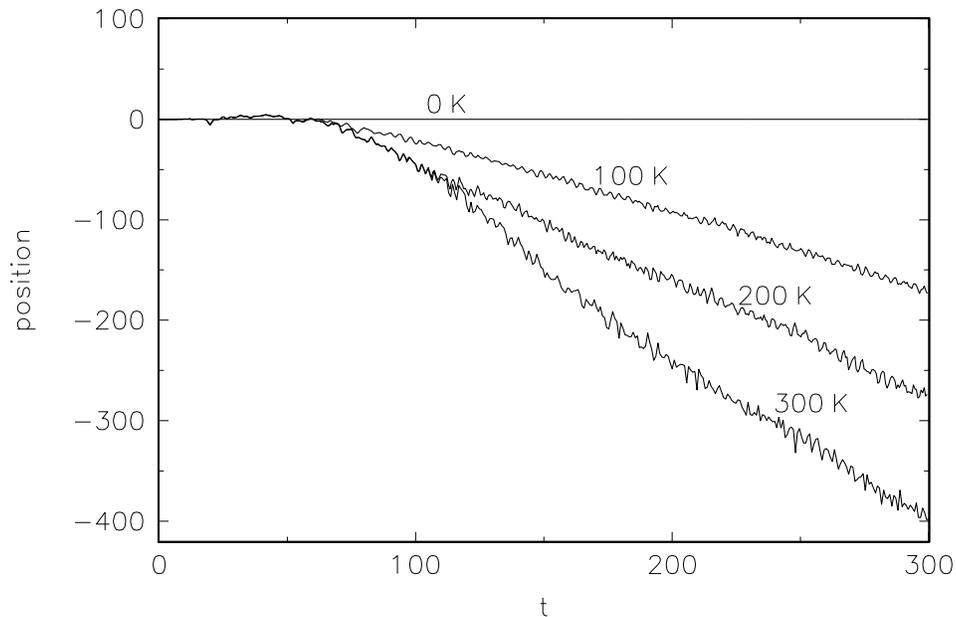}\\
\caption{Influence of temperature on the time-evolution of the
breather center. Temperatures are indicated on the curves.}
 \label{Fig23}
\end{figure}

This quantity describes the temporal behavior of the position of
the center of a breather. Thus it represents a  measure for the
mobility of the breathers. Generally, the higher the temperature
the larger the amplitude of the radial breather becomes and the
faster the radial (and torsional) breather travels along the DNA
lattice. This behavior is illustrated in Fig.~\ref{Fig23}.


In summary, out of an initial non-equilibrium situation, for which
the hydrogen bonds of an under-twisted segment of the DNA lattice
have been stretched, breathers develop in the radial and angular
displacement variables. For zero temperature the breathers remain
standing in the initially excited region. However, for $T>0$ the
breathers start to travel coherently towards an end of the DNA
duplex. The observed breathers represent realistically the
oscillating bubbles found prior to complete unzipping of DNA, i.e.
they oscillate with periods in the range $0.3\,-\,0.8\,ps$, are on
the average extended over $10-20$ base pairs and possess maximal
amplitudes of the order of $\lesssim 0.6$\,\AA. These results
demonstrate that the action of some enzyme, mimicked by localized
radial and torsional distortions of the DNA equilibrium
configuration, initiates in fact the production of oscillating
bubbles in DNA. Moreover, these "oscillating bubbles" sustain the
impact of thermal perturbations.

\section*{Acknowledgments}
We thank all the colleagues with whom we had very fruitful
collaborations and discussions over the years. Without them this
work would not have been possible.
 This work has been supported by the MECD--FEDER project FIS2004-01183.

\newcommand{\noopsort}[1]{} \newcommand{\printfirst}[2]{#1}
  \newcommand{\singleletter}[1]{#1} \newcommand{\switchargs}[2]{#2#1}

\end{document}